\input harvmac

\input amssym

%\draftmode

\def\IZ{\Bbb{Z}}

\def\IR{\Bbb{R}}

\def\p{\partial}

\def\taub{\overline{\tau}}
\def\Lt{\tilde{L}}
\def\wb{\overline{w}}

\def\Jt{\tilde{J}}

\def\qb{\overline{q}}
\def\zt{\tilde{z}}

\def\kt{\tilde{k}}

\def\mut{\tilde{\mu}}
\def\chit{\tilde{\chi}}

\def\mut{\tilde{\mu}}
\def\qt{\tilde{q}}

\def\hbar{\overline{h}}

\def\At{\tilde{A}}
\def\kt{\tilde{k}}
\def\ct{\tilde{c}}

%%%%%%%%%%%%%%%%%%%%%%%%%%%%%%%%%%%%%%%%%%%%%%
%%%%%%%%%%%%%%%%%%%%%%%%%%%
% Title
%%%%%%%%%%%%%%%%%%%%%%%%%%%%%%%%%%%%%%%%%%%%%%
%%%%%%%%%%%%%%%%%%%%%%%%%%%

\Title{\vbox{\baselineskip12pt
%\hbox{hep-th/0508218}
%\hbox{UCLA-05-TEP-XX} \hbox{MCTP-XX-XX}
}} {\vbox{\centerline {Partition functions and elliptic genera from
supergravity}}}
%\medskip\vbox{\centerline {in Theories with Higher
%}}} }
\centerline{Per
Kraus\foot{pkraus@physics.ucla.edu} and Finn
Larsen\foot{larsenf@umich.edu}}

\bigskip
\centerline{${}^1$\it{Department of Physics and Astronomy,
UCLA,}}\centerline{\it{ Los Angeles, CA 90095-1547,
USA.}}\vskip.2cm \centerline{${}^2$\it{Michigan Center for
Theoretical Physics, Department of Physics}} \centerline{\it{University of Michigan, Ann
Arbor, MI 48109-1120, USA.}}

\baselineskip15pt

\vskip .3in

\centerline{\bf Abstract}  We develop the spacetime aspects of the computation of
partition functions  for string/M-theory on AdS$_3 \times M$.    Subleading corrections
to the semi-classical result are included systematically, laying the groundwork for comparison with  CFT partition functions  via the AdS$_3 $/CFT$_2$ correspondence. 
This leads to a better  understanding of the ``Farey tail" expansion of Dijkgraaf
et. al.  from the point of view of bulk physics.    Besides clarifying various issues,
we also extend the analysis to the ${\cal N}=2$ setting with higher derivative effects included. 

%%%
\Date{July, 2006}
%%%%%%%%%%%%%%%%%%%%%%%%%%%%%%%%%%%%%%%%%%%%%%

%\GaiottoNS
\lref\GaiottoNS{
  D.~Gaiotto, A.~Strominger and X.~Yin,
  ``From AdS(3)/CFT(2) to black holes / topological strings,''
  arXiv:hep-th/0602046.
  %%CITATION = HEP-TH 0602046;%%
}

%\GaiottoWM
\lref\GaiottoWM{
  D.~Gaiotto, A.~Strominger and X.~Yin,
  ``The M5-brane elliptic genus: Modularity and BPS states,''
  arXiv:hep-th/0607010.
  %%CITATION = HEP-TH 0607010;%%
}

%\deBoerIP
\lref\deBoerIP{
  F.~Larsen,
  ``The perturbation spectrum of black holes in N = 8 supergravity,''
  Nucl.\ Phys.\ B {\bf 536}, 258 (1998)
  [arXiv:hep-th/9805208].
  %%CITATION = HEP-TH 9805208;%%
  J.~de Boer,
  ``Six-dimensional supergravity on S**3 x AdS(3) and 2d conformal field
  theory,''
  Nucl.\ Phys.\ B {\bf 548}, 139 (1999)
  [arXiv:hep-th/9806104];
  %%CITATION = HEP-TH 9806104;%%
}
%\FujiiTC
\lref\FujiiTC{
  A.~Fujii, R.~Kemmoku and S.~Mizoguchi,
  %``D = 5 simple supergravity on AdS(3) x S(2) and N = 4 superconformal  field
  %theory,''
  Nucl.\ Phys.\ B {\bf 574}, 691 (2000)
  [arXiv:hep-th/9811147].
  %%CITATION = HEP-TH 9811147;%%
}

%\DijkgraafFQ
\lref\farey{
  R.~Dijkgraaf, J.~M.~Maldacena, G.~W.~Moore and E.~P.~Verlinde,
  ``A black hole farey tail,''
  arXiv:hep-th/0005003.
  %%CITATION = HEP-TH 0005003;%%
}

%\PolchinskiRQ
\lref\PolchinskiRQ{
  J.~Polchinski,
  ``String theory. Vol. 1: An introduction to the bosonic string,''
%\href{http://www.slac.stanford.edu/spires/find/hep/www?irn=4634799}{SPIRES entry}
}

%\KrausVZ
\lref\KrausVZ{
  P.~Kraus and F.~Larsen,
  ``Microscopic black hole entropy in theories with higher derivatives,''
  JHEP {\bf 0509}, 034 (2005)
  [arXiv:hep-th/0506176];
  %%CITATION = HEP-TH 0506176;%%
%\KrausZM
  ``Holographic gravitational anomalies,''
  JHEP {\bf 0601}, 022 (2006)
  [arXiv:hep-th/0508218].
  %%CITATION = HEP-TH 0508218;%%
}

%\MaldacenaDR
\lref\MaldacenaDR{
  J.~M.~Maldacena and L.~Maoz,
  ``De-singularization by rotation,''
  JHEP {\bf 0212}, 055 (2002)
  [arXiv:hep-th/0012025];
  O.~Lunin, J.~M.~Maldacena and L.~Maoz,
  ``Gravity solutions for the D1-D5 system with angular momentum,''
  arXiv:hep-th/0212210;
  %%CITATION = HEP-TH 0212210;%%
  }

  %\GopakumarII
\lref\GopakumarII{
  R.~Gopakumar and C.~Vafa,
  ``M-theory and topological strings. I,''
  arXiv:hep-th/9809187.  ``M-theory and topological strings. II,''
  arXiv:hep-th/9812127.
  %%CITATION = HEP-TH 9812127;%%
  %%CITATION = HEP-TH 9809187;%%
}

  %\KatzXQ
\lref\KatzXQ{
  S.~Katz, A.~Klemm and C.~Vafa,
  ``M-theory, topological strings and spinning black holes,''
  Adv.\ Theor.\ Math.\ Phys.\  {\bf 3}, 1445 (1999)
  [arXiv:hep-th/9910181].
  %%CITATION = HEP-TH 9910181;%%
}

%\LuninBJ
\lref\LuninBJ{
  O.~Lunin, S.~D.~Mathur and A.~Saxena,
  ``What is the gravity dual of a chiral primary?,''
  Nucl.\ Phys.\ B {\bf 655}, 185 (2003)
  [arXiv:hep-th/0211292].
  %%CITATION = HEP-TH 0211292;%%
}

%\MathurZP
\lref\MathurZP{
  S.~D.~Mathur,
  ``The fuzzball proposal for black holes: An elementary review,''
  Fortsch.\ Phys.\  {\bf 53}, 793 (2005)
  [arXiv:hep-th/0502050].
  %%CITATION = HEP-TH 0502050;%%
}

%\BalasubramanianRT
\lref\BalasubramanianRT{
  V.~Balasubramanian, J.~de Boer, E.~Keski-Vakkuri and S.~F.~Ross,
  ``Supersymmetric conical defects: Towards a string theoretic description  of
  black hole formation,''
  Phys.\ Rev.\ D {\bf 64}, 064011 (2001)
  [arXiv:hep-th/0011217].
  %%CITATION = HEP-TH 0011217;%%
}

%\KutasovZH
\lref\KutasovZH{
  D.~Kutasov, F.~Larsen and R.~G.~Leigh,
  ``String theory in magnetic monopole backgrounds,''
  Nucl.\ Phys.\ B {\bf 550}, 183 (1999)
  [arXiv:hep-th/9812027].
  %%CITATION = HEP-TH 9812027;%%
}
%\CveticJA
\lref\CveticJA{
  M.~Cvetic and F.~Larsen,
  ``Statistical entropy of four-dimensional rotating black holes from
  near-horizon geometry,''
  Phys.\ Rev.\ Lett.\  {\bf 82}, 484 (1999)
  [arXiv:hep-th/9805146].
  %%CITATION = HEP-TH 9805146;%%
}
%\CveticXH
\lref\CveticXH{
  M.~Cvetic and F.~Larsen,
  ``Near horizon geometry of rotating black holes in five dimensions,''
  Nucl.\ Phys.\ B {\bf 531}, 239 (1998)
  [arXiv:hep-th/9805097].
  %%CITATION = HEP-TH 9805097;%%
}

\lref\fareytale{J.~M.~Maldacena and A.~Strominger,
  ``AdS(3) black holes and a stringy exclusion principle,''
  JHEP {\bf 9812}, 005 (1998)
  [arXiv:hep-th/9804085];
  R.~Dijkgraaf, J.~M.~Maldacena, G.~W.~Moore and E.~Verlinde,
  ``A black hole farey tail,''
  arXiv:hep-th/0005003.
  %%CITATION = HEP-TH 0005003;%%

  %%CITATION = HEP-TH 9804085;%%
}

\lref\generalRRRR{
%gutperle, duffliu
  M.~B.~Green and J.~H.~Schwarz,
  ``Supersymmetrical Dual String Theory. 2. Vertices And Trees,''
  Nucl.\ Phys.\ B {\bf 198}, 252 (1982);
  %%CITATION = NUPHA,B198,252;%%
  D.~J.~Gross and E.~Witten,
  ``Superstring Modifications Of Einstein's Equations,''
  Nucl.\ Phys.\ B {\bf 277}, 1 (1986);
  %%CITATION = NUPHA,B277,1;%%
  W.~Lerche, B.~E.~W.~Nilsson and A.~N.~Schellekens,
  ``Heterotic String Loop Calculation Of The Anomaly Cancelling Term,''
  Nucl.\ Phys.\ B {\bf 289}, 609 (1987);
  %%CITATION = NUPHA,B289,609;%%
  M.~J.~Duff, J.~T.~Liu and R.~Minasian,
  ``Eleven-dimensional origin of string / string duality: A one-loop test,''
  Nucl.\ Phys.\ B {\bf 452}, 261 (1995)
  [arXiv:hep-th/9506126];
  %%CITATION = HEP-TH 9506126;%%
   M.~B.~Green, M.~Gutperle and P.~Vanhove,
  ``One loop in eleven dimensions,''
  Phys.\ Lett.\ B {\bf 409}, 177 (1997)
  [arXiv:hep-th/9706175];
  %%CITATION = HEP-TH 9706175;%%
    J.~G.~Russo and A.~A.~Tseytlin,
  ``One-loop four-graviton amplitude in eleven-dimensional supergravity,''
  Nucl.\ Phys.\ B {\bf 508}, 245 (1997)
  [arXiv:hep-th/9707134]; P.~S.~Howe and D.~Tsimpis,
  ``On higher-order corrections in M theory,''
  JHEP {\bf 0309}, 038 (2003)
  [arXiv:hep-th/0305129].
  %%CITATION = HEP-TH 9707134;%%
}
%also dewitt+behrndt

\lref\masakiref{
%\IizukaUV
  N.~Iizuka and M.~Shigemori,
  ``A Note on D1-D5-J System and 5D Small Black Ring,''
  arXiv:hep-th/0506215.
  %%CITATION = HEP-TH 0506215;%%
}

\lref\russuss{  J.~G.~Russo and L.~Susskind,
  ``Asymptotic level density in heterotic string theory and rotating black
  holes,''
  Nucl.\ Phys.\ B {\bf 437}, 611 (1995)
  [arXiv:hep-th/9405117].
  %%CITATION = HEP-TH 9405117;%%
  }

  %\KrausGH
\lref\KLatt{
  P.~Kraus and F.~Larsen,
  ``Attractors and black rings,''
  Phys.\ Rev.\ D {\bf 72}, 024010 (2005)
  [arXiv:hep-th/0503219].
  %%CITATION = HEP-TH 0503219;%%
}

%\ChamseddinePI
\lref\ChamseddinePI{
  A.~H.~Chamseddine, S.~Ferrara, G.~W.~Gibbons and R.~Kallosh,
  ``Enhancement of supersymmetry near 5d black hole horizon,''
  Phys.\ Rev.\ D {\bf 55}, 3647 (1997)
  [arXiv:hep-th/9610155].
  %%CITATION = HEP-TH 9610155;%%
}

\lref\supertube{D.~Mateos and P.~K.~Townsend,
  ``Supertubes,''
  Phys.\ Rev.\ Lett.\  {\bf 87}, 011602 (2001)
  [arXiv:hep-th/0103030];
  %%CITATION = HEP-TH 0103030;%%
  R.~Emparan, D.~Mateos and P.~K.~Townsend,
  ``Supergravity supertubes,''
  JHEP {\bf 0107}, 011 (2001)
  [arXiv:hep-th/0106012].
  %%CITATION = HEP-TH 0106012;%%
}

\lref\Mathurrev{ S.~D.~Mathur,``The fuzzball proposal for black
holes: An elementary review,'' arXiv:hep-th/0502050.
  %%CITATION = HEP-TH 0502050;%%
}

\lref\StromBTZ{ A.~Strominger,
 ``Black hole entropy from near-horizon microstates'',
JHEP {\bf 9802}, 009 (1998); [arXiv:hep-th/9712251];
V.~Balasubramanian and F.~Larsen, ``Near horizon geometry and
black holes in four dimensions'', Nucl.\ Phys.\ B {\bf 528}, 229
(1998); [arXiv:hep-th/9802198].
%%CITATION = HEP-TH 9802198;%%
}

\lref\MSW{ J.~M.~Maldacena, A.~Strominger and E.~Witten, ``Black
hole entropy in M-theory'', JHEP {\bf 9712}, 002 (1997);
[arXiv:hep-th/9711053]. }
  %%CITATION = HEP-TH 9711053;%%

\lref\HMM{ J.~A.~Harvey, R.~Minasian and G.~W.~Moore,
``Non-abelian tensor-multiplet anomalies,''
 JHEP {\bf 9809}, 004 (1998)
  [arXiv:hep-th/9808060].
 %%CITATION = HEP-TH 9808060;%%
}

\lref\tseytRRRR{  A.~A.~Tseytlin,
  ``R**4 terms in 11 dimensions and conformal anomaly of (2,0) theory,''
  Nucl.\ Phys.\ B {\bf 584}, 233 (2000)
  [arXiv:hep-th/0005072].
  %%CITATION = HEP-TH 0005072;%%
}

\lref\antRRRR{  I.~Antoniadis, S.~Ferrara, R.~Minasian and
K.~S.~Narain,
  ``R**4 couplings in M- and type II theories on Calabi-Yau spaces,''
  Nucl.\ Phys.\ B {\bf 507}, 571 (1997)
  [arXiv:hep-th/9707013].
  %%CITATION = HEP-TH 9707013;%%
 }

\lref\WittenMfive{ E.~Witten,
  ``Five-brane effective action in M-theory,''
  J.\ Geom.\ Phys.\  {\bf 22}, 103 (1997)
  [arXiv:hep-th/9610234].
  %%CITATION = HEP-TH 9610234;%%
}

\lref\wittenAdS{ E.~Witten,
  ``Anti-de Sitter space and holography,''
  Adv.\ Theor.\ Math.\ Phys.\  {\bf 2}, 253 (1998)
  [arXiv:hep-th/9802150].
  %%CITATION = HEP-TH 9802150;%%
  }

\lref\iosef{  I.~Bena,
  ``Splitting hairs of the three charge black hole,''
  Phys.\ Rev.\ D {\bf 70}, 105018 (2004)
  [arXiv:hep-th/0404073].
  %%CITATION = HEP-TH 0404073;%%
}

\lref\brownhen{  J.~D.~Brown and M.~Henneaux,
 ``Central Charges In The Canonical Realization Of Asymptotic Symmetries: An
  Example From Three-Dimensional Gravity,''
  Commun.\ Math.\ Phys.\  {\bf 104}, 207 (1986).
  %%CITATION = CMPHA,104,207;%%
  }

\lref\wald{
  R.~M.~Wald,
  ``Black hole entropy is the Noether charge,''
  Phys.\ Rev.\ D {\bf 48}, 3427 (1993)
  [arXiv:gr-qc/9307038].
  %%CITATION = GR-QC 9307038;%%
R.~Wald, Phys.\ Rev.\ D {\bf 48} R3427 (1993);
   V.~Iyer and R.~M.~Wald,
  ``Some properties of Noether charge and a proposal for dynamical black hole
  entropy,''
  Phys.\ Rev.\ D {\bf 50}, 846 (1994)
  [arXiv:gr-qc/9403028].
  %%CITATION = GR-QC 9403028;%%
 ``A Comparison of Noether charge and Euclidean methods for computing the
  entropy of stationary black holes,''
  Phys.\ Rev.\ D {\bf 52}, 4430 (1995)
  [arXiv:gr-qc/9503052].
  %%CITATION = GR-QC 9503052;%%
}

\lref\Sen{  A.~Sen,
  ``How does a fundamental string stretch its horizon?,''
  JHEP {\bf 0505}, 059 (2005)
  [arXiv:hep-th/0411255];
  %%CITATION = HEP-TH 0411255;%%
   ``Black holes, elementary strings and holomorphic anomaly,''
  arXiv:hep-th/0502126;
  %%CITATION = HEP-TH 0502126;%%;
   ``Stretching the horizon of a higher dimensional small black hole,''
  arXiv:hep-th/0505122; ``Black hole entropy function and the attractor mechanism in higher
  derivative gravity,''
  arXiv:hep-th/0506177; ``Entropy Function for Heterotic Black Holes,''
  arXiv:hep-th/0508042.
  %%CITATION = HEP-TH 0505122;%%
  }

\lref\saidasoda{
  H.~Saida and J.~Soda,
  ``Statistical entropy of BTZ black hole in higher curvature gravity,''
  Phys.\ Lett.\ B {\bf 471}, 358 (2000)
  [arXiv:gr-qc/9909061].
  %%CITATION = GR-QC 9909061;%%
}

\lref\attract{ S.~Ferrara, R.~Kallosh and A.~Strominger, ``N=2
extremal black holes'', Phys.\ Rev.\ D {\bf 52}, 5412 (1995),
[arXiv:hep-th/9508072];
  %%CITATION = HEP-TH 9508072;%%
 A.~Strominger,
 ``Macroscopic Entropy of $N=2$ Extremal Black Holes'',
 Phys.\ Lett.\ B {\bf 383}, 39 (1996),
[arXiv:hep-th/9602111];
  %%CITATION = HEP-TH 9602111;%%
S.~Ferrara and R.~Kallosh, ``Supersymmetry and Attractors'',
Phys.\ Rev.\ D {\bf 54}, 1514 (1996), [arXiv:hep-th/9602136];
  %%CITATION = HEP-TH 9602136;%%
``Universality of Supersymmetric Attractors'', Phys.\ Rev.\ D {\bf
54}, 1525 (1996), [arXiv:hep-th/9603090];
 %%CITATION = HEP-TH 9603090;%%
R.~Kallosh, A.~Rajaraman and W.~K.~Wong, ``Supersymmetric rotating
black holes and attractors'', Phys.\ Rev.\ D {\bf 55}, 3246
(1997), [arXiv:hep-th/9611094];
  %%CITATION = HEP-TH 9611094;%%
A~Chou, R.~Kallosh, J.~Rahmfeld, S.~J.~Rey, M.~Shmakova and
W.~K.~Wong, ``Critical points and phase transitions in 5d
compactifications of M-theory''. Nucl.\ Phys.\ B {\bf 508}, 147
(1997); [arXiv:hep-th/9704142].
 %%CITATION = HEP-TH 9704142;%%
}

\lref\moore{G.~W.~Moore,``Attractors and arithmetic'',
[arXiv:hep-th/9807056];
%%CITATION = HEP-TH 9807056;%%
``Arithmetic and attractors'', [arXiv:hep-th/9807087];
  %%CITATION = HEP-TH 9807087;%%
``Les Houches lectures on strings and arithmetic'',
[arXiv:hep-th/0401049];
  %%CITATION = HEP-TH 0401049;%%
B.~R.~Greene and C.~I.~Lazaroiu, ``Collapsing D-branes in
Calabi-Yau moduli space. I'', Nucl.\ Phys.\ B {\bf 604}, 181
(2001), [arXiv:hep-th/0001025]. }

%\ChamseddinePI
\lref\ChamseddinePI{
  A.~H.~Chamseddine, S.~Ferrara, G.~W.~Gibbons and R.~Kallosh,
  ``Enhancement of supersymmetry near 5d black hole horizon,''
  Phys.\ Rev.\ D {\bf 55}, 3647 (1997)
  [arXiv:hep-th/9610155].
  %%CITATION = HEP-TH 9610155;%%
}

\lref\denef{  %%CITATION = HEP-TH 0001025;%%
F.~Denef,``Supergravity flows and D-brane stability'', JHEP {\bf
0008}, 050 (2000), [arXiv:hep-th/0005049];
%%CITATION = HEP-TH 0005049;%%
``On the correspondence between D-branes and stationary
supergravity
 solutions of type II Calabi-Yau compactifications'',
[arXiv:hep-th/0010222];
 %%CITATION = HEP-TH 0010222;%%
``(Dis)assembling special Lagrangians'', [arXiv:hep-th/0107152].
%%CITATION = HEP-TH 0107152;%%
  B.~Bates and F.~Denef,
   ``Exact solutions for supersymmetric stationary black hole composites,''
  arXiv:hep-th/0304094.
}

\lref\OSV{H.~Ooguri, A.~Strominger and C.~Vafa, ``Black hole
attractors and the topological string'', Phys.\ Rev.\ D {\bf 70},
106007 (2004), [arXiv:hep-th/0405146].
%%CITATION = HEP-TH 0405146;%%
}

 \lref\VerlindeHolAn{

E.~Verlinde, ``Attractors and the holomorphic anomaly'',
[arXiv:hep-th/0412139];
  %%CITATION = HEP-TH 0412139;%%
 }

 %\GaiottoNS
\lref\GSY{
  D.~Gaiotto, A.~Strominger and X.~Yin,
  ``From AdS(3)/CFT(2) to black holes / topological strings,''
  arXiv:hep-th/0602046.
  %%CITATION = HEP-TH 0602046;%%
}

%\BrownNW
\lref\BrownNW{
  J.~D.~Brown and M.~Henneaux,
  ``Central Charges In The Canonical Realization Of Asymptotic Symmetries: An
  Example From Three-Dimensional Gravity,''
  Commun.\ Math.\ Phys.\  {\bf 104}, 207 (1986).
  %%CITATION = CMPHA,104,207;%%
}

 \lref\DDMP{
 A.~Dabholkar, F.~Denef, G.~W.~Moore and B.~Pioline,
  ``Exact and asymptotic degeneracies of small black holes,''
  JHEP {\bf 0508}, 021 (2005)
  [arXiv:hep-th/0502157];
  %%CITATION = HEP-TH 0502157;%%
   ``Precision counting of small black holes,''
  JHEP {\bf 0510}, 096 (2005)
  [arXiv:hep-th/0507014].
  %%CITATION = HEP-TH 0507014;%%
  %%CITATION = HEP-TH 0502157;%%
}

\lref\curvcorr{A.~Dabholkar, ``Exact counting of black hole
microstates", [arXiv:hep-th/0409148],
%%CITATION = HEP-TH 0409148;%%
A.~Dabholkar, R.~Kallosh and A.~Maloney, ``A stringy cloak for a
classical singularity'', JHEP {\bf 0412}, 059 (2004),
[arXiv:hep-th/0410076].
 %%CITATION = HEP-TH 0410076;%%
}

 \lref\bkmicro{
 I.~Bena and P.~Kraus,
 ``Microscopic description of black rings in AdS/CFT'',
JHEP {\bf 0412}, 070 (2004)
  [arXiv:hep-th/0408186].
%%CITATION = HEP-TH 0408186;%%
} \lref\cgms{ M.~Cyrier, M.~Guica, D.~Mateos and A.~Strominger,
``Microscopic entropy of the black ring'', [arXiv:hep-th/0411187].
  %%CITATION = HEP-TH 0411187;%%
}

%\GunaydinBI
\lref\GunaydinBI{
  M.~Gunaydin, G.~Sierra and P.~K.~Townsend,
  ``The Geometry Of N=2 Maxwell-Einstein Supergravity And Jordan
  Algebras'',
  Nucl.\ Phys.\ B {\bf 242}, 244 (1984);
  ``Gauging The D = 5 Maxwell-Einstein Supergravity Theories:
   More On Jordan Algebras,''
  Nucl.\ Phys.\ B {\bf 253}, 573 (1985).
  %%CITATION = NUPHA,B253,573;%%
}

%\deWitCR
\lref\deWitCR{
  B.~de Wit and A.~Van Proeyen,
  ``Broken sigma model isometries in very special geometry,''
  Phys.\ Lett.\ B {\bf 293}, 94 (1992)
  [arXiv:hep-th/9207091].
  %%CITATION = HEP-TH 9207091;%%
}

%\CadavidBK
\lref\CadavidBK{
  A.~C.~Cadavid, A.~Ceresole, R.~D'Auria and S.~Ferrara,
  ``Eleven-dimensional supergravity compactified on Calabi-Yau threefolds,''
  Phys.\ Lett.\ B {\bf 357}, 76 (1995)
  [arXiv:hep-th/9506144].
  %%CITATION = HEP-TH 9506144;%%
}

%\PapadopoulosDA
\lref\PapadopoulosDA{
  G.~Papadopoulos and P.~K.~Townsend,
  ``Compactification of D = 11 supergravity on spaces of exceptional
  holonomy,''
  Phys.\ Lett.\ B {\bf 357}, 300 (1995)
  [arXiv:hep-th/9506150].
  %%CITATION = HEP-TH 9506150;%%
}

%\AntoniadisCY
\lref\AntoniadisCY{
  I.~Antoniadis, S.~Ferrara and T.~R.~Taylor,
  ``N=2 Heterotic Superstring and its Dual Theory in Five Dimensions,''
  Nucl.\ Phys.\ B {\bf 460}, 489 (1996)
  [arXiv:hep-th/9511108].
  %%CITATION = HEP-TH 9511108;%%
}

%\GauntlettNW
\lref\GauntlettNW{
  J.~P.~Gauntlett, J.~B.~Gutowski, C.~M.~Hull, S.~Pakis and H.~S.~Reall,
  ``All supersymmetric solutions of minimal supergravity in five dimensions,''
  Class.\ Quant.\ Grav.\  {\bf 20}, 4587 (2003)
  [arXiv:hep-th/0209114]}
  %%CITATION = HEP-TH 0209114;%%}

\lref\gutow{
  J.~B.~Gutowski and H.~S.~Reall,
  ``General supersymmetric AdS(5) black holes'',
  JHEP {\bf 0404}, 048 (2004)
  [arXiv:hep-th/0401129];
  J.~B.~Gutowski,
  ``Uniqueness of five-dimensional supersymmetric black holes'',
  JHEP {\bf 0408}, 049 (2004)
  [arXiv:hep-th/0404079].
}

%\BenaDE
\lref\BenaDE{
  I.~Bena and N.~P.~Warner,
  ``One ring to rule them all ... and in the darkness bind them?'',
  [arXiv:hep-th/0408106].
  %%CITATION = HEP-TH 0408106;%%
}
%\BMPV
\lref\BMPV{ J.~C.~Breckenridge, R.~C.~Myers, A.~W.~Peet and
C.~Vafa, ``D-branes and spinning black holes'', Phys.\ Lett.\ B
{\bf 391}, 93 (1997); [arXiv:hep-th/9602065].
  %%CITATION = HEP-TH 9602065;%%
}

\lref\EEMR{H.~Elvang, R.~Emparan, D.~Mateos and H.~S.~Reall,
``Supersymmetric black rings and three-charge supertubes'',
  Phys.\ Rev.\ D {\bf 71}, 024033 (2005);
  [arXiv:hep-th/0408120].
  %%CITATION = HEP-TH 0408120;%%
}

%\ElvangRT
\lref\ElvangRT{
  H.~Elvang, R.~Emparan, D.~Mateos and H.~S.~Reall,
  ``A supersymmetric black ring'',
  Phys.\ Rev.\ Lett.\  {\bf 93}, 211302 (2004)
  [arXiv:hep-th/0407065].
  %%CITATION = HEP-TH 0407065;%%
}

%\BenaWT
\lref\BenaWT{
  I.~Bena and P.~Kraus,
  ``Three charge supertubes and black hole hair,''
  Phys.\ Rev.\ D {\bf 70}, 046003 (2004)
  [arXiv:hep-th/0402144].
  %%CITATION = HEP-TH 0402144;%%
}

%\GauntlettQY
\lref\GauntlettQY{ J.~P.~Gauntlett and J.~B.~Gutowski, ``General
Concentric  Black Rings'', [arXiv:hep-th/0408122]. J.~P.~Gauntlett
and J.~B.~Gutowski, ``Concentric  black rings'',
[arXiv:hep-th/0408010].
%%CITATION = HEP-TH 0408122;%%
}

%\DenefNB
\lref\denefa{
  F.~Denef,
   ``Supergravity flows and D-brane stability'',
JHEP {\bf 0008}, 050 (2000), [arXiv:hep-th/0005049].
  %%CITATION = HEP-TH 0005049;%%
}

%\BatesVX
\lref\denefc{
  B.~Bates and F.~Denef,
``Exact solutions for supersymmetric stationary black hole
composites'', [arXiv:hep-th/0304094].
  %%CITATION = HEP-TH 0304094;%%
}

%\SenPU
\lref\SenPU{ A.~Sen, ``Black holes, elementary strings and
holomorphic anomaly'',
 [arXiv:hep-th/0502126].
%%CITATION = HEP-TH 0502126;%%
}

%\CardosoFP
\lref\CardosoFP{ 

  G.~Lopes Cardoso, B.~de Wit and T.~Mohaupt,
   ``Macroscopic entropy formulae and non-holomorphic corrections for
  supersymmetric black holes'',
  Nucl.\ Phys.\ B {\bf 567}, 87 (2000)
  [arXiv:hep-th/9906094];
  ``Deviations from the area law for supersymmetric black holes'',
  Fortsch.\ Phys.\  {\bf 48}, 49 (2000)
  [arXiv:hep-th/9904005];
  ``Corrections to macroscopic supersymmetric black-hole entropy'',
  Phys.\ Lett.\ B {\bf 451}, 309 (1999)
  [arXiv:hep-th/9812082].
  %%CITATION = HEP-TH 9812082;%%
}

%\BenaTD
\lref\BenaTD{
  I.~Bena, C.~W.~Wang and N.~P.~Warner,
  ``Black rings with varying charge density'',
[arXiv:hep-th/0411072].
  %%CITATION = HEP-TH 0411072;%%
}

%\BenaWV
\lref\BenaWV{ I.~Bena, ``Splitting hairs of the three charge black
hole'', Phys.\ Rev.\ D {\bf 70}, 105018 (2004),
[arXiv:hep-th/0404073].
  %%CITATION = HEP-TH 0404073;%%
}

\lref\rings{ H.~Elvang, R.~Emparan, D.~Mateos and H.~S.~Reall, ``A
supersymmetric black ring,''  Phys. Rev. Lett.  {\bf 93}, 211302
(2004) [arXiv:hep-th/0407065]; ``Supersymmetric black rings and
three-charge supertubes,'' Phys. Rev. D {\bf 71}, 024033 (2005)
[arXiv:hep-th/0408120]; I.~Bena and N.~P.~Warner, ``One ring to
rule them all ... and in the darkness bind them?,''
arXiv:hep-th/0408106; J.~P.~Gauntlett and J.~B.~Gutowski,
``General concentric black rings,'' Phys.\ Rev. D {\bf 71}, 045002
(2005) [arXiv:hep-th/0408122].}

\lref\hensken{  M.~Henningson and K.~Skenderis,
  ``The holographic Weyl anomaly,''
  JHEP {\bf 9807}, 023 (1998)
  [arXiv:hep-th/9806087].
  %%CITATION = HEP-TH 9806087;%%
  }

\lref\balkraus{  V.~Balasubramanian and P.~Kraus,
  ``A stress tensor for anti-de Sitter gravity,''
  Commun.\ Math.\ Phys.\  {\bf 208}, 413 (1999)
  [arXiv:hep-th/9902121].
  %%CITATION = HEP-TH 9902121;%%
  }

  %\MarolfFY
\lref\MarolfFY{
  D.~Marolf and B.~C.~Palmer,
  ``Gyrating strings: A new instability of black strings?,''
  Phys.\ Rev.\ D {\bf 70}, 084045 (2004)
  [arXiv:hep-th/0404139].
  %%CITATION = HEP-TH 0404139;%%
}

%\MyersPS
\lref\MyersPS{
  R.~C.~Myers,
  ``Stress tensors and Casimir energies in the AdS/CFT correspondence,''
  Phys.\ Rev.\ D {\bf 60}, 046002 (1999)
  [arXiv:hep-th/9903203].
  %%CITATION = HEP-TH 9903203;%%
}

%\KrausDI
\lref\KrausDI{
  P.~Kraus, F.~Larsen and R.~Siebelink,
  ``The gravitational action in asymptotically AdS and flat spacetimes,''
  Nucl.\ Phys.\ B {\bf 563}, 259 (1999)
  [arXiv:hep-th/9906127].
  %%CITATION = HEP-TH 9906127;%%
}

%\deHaroXN
\lref\deHaroXN{
  S.~de Haro, S.~N.~Solodukhin and K.~Skenderis,
  ``Holographic reconstruction of spacetime and renormalization in the  AdS/CFT
  correspondence,''
  Commun.\ Math.\ Phys.\  {\bf 217}, 595 (2001)
  [arXiv:hep-th/0002230].
  %%CITATION = HEP-TH 0002230;%%
}

%\PapadimitriouII
\lref\PapadimitriouII{
  I.~Papadimitriou and K.~Skenderis,
  ``Thermodynamics of asymptotically locally AdS spacetimes,''
  arXiv:hep-th/0505190.
  %%CITATION = HEP-TH 0505190;%%
}

%\HollandsWT
\lref\HollandsWT{
  S.~Hollands, A.~Ishibashi and D.~Marolf,
  ``Comparison between various notions of conserved charges in asymptotically
  AdS-spacetimes,''
  Class.\ Quant.\ Grav.\  {\bf 22}, 2881 (2005)
  [arXiv:hep-th/0503045]; ``Counter-term charges generate bulk symmetries,''
  arXiv:hep-th/0503105.
  %%CITATION = HEP-TH 0503045;%%
}

%\AlvarezGaumeIG
\lref\AlvarezGaumeIG{
  L.~Alvarez-Gaume and E.~Witten,
  ``Gravitational Anomalies,''
  Nucl.\ Phys.\ B {\bf 234}, 269 (1984).
  %%CITATION = NUPHA,B234,269;%%
}

%\GinspargQN
\lref\GinspargQN{
  P.~H.~Ginsparg,
  ``Applications Of Topological And Differential Geometric Methods To Anomalies
  In Quantum Field Theory,''
HUTP-85/A056
%\href{http://www.slac.stanford.edu/spires/find/hep/www?r=hutp-85\%2Fa056}{SPIRES entry}
{\it To appear in Proc. of 16th GIFT Seminar on Theoretical
Physics, Jaca, Spain, Jun 3-7, 1985} }

%\StromingerYG
\lref\StromingerYG{
  A.~Strominger,
  ``AdS(2) quantum gravity and string theory,''
  JHEP {\bf 9901}, 007 (1999)
  [arXiv:hep-th/9809027].
  %%CITATION = HEP-TH 9809027;%%
}

\lref\mtw{C.W. Misner, K.~Thorne, and J.A.~Wheeler,
``Gravitation,"  W. H. Freeman, (1973).}

%\BardeenPM
\lref\BardeenPM{
  W.~A.~Bardeen and B.~Zumino,
  ``Consistent And Covariant Anomalies In Gauge And Gravitational Theories,''
  Nucl.\ Phys.\ B {\bf 244}, 421 (1984).
  %%CITATION = NUPHA,B244,421;%%
}

%\AlvarezGaumeDR
\lref\AlvarezGaumeDR{
  L.~Alvarez-Gaume and P.~H.~Ginsparg,
  ``The Structure Of Gauge And Gravitational Anomalies,''
  Annals Phys.\  {\bf 161}, 423 (1985)
  [Erratum-ibid.\  {\bf 171}, 233 (1986)].
  %%CITATION = APNYA,161,423;%%
}

%\BrownBR
\lref\BrownBR{
  J.~D.~Brown and J.~W.~.~York,
  ``Quasilocal energy and conserved charges derived from the gravitational
  %action,''
  Phys.\ Rev.\ D {\bf 47}, 1407 (1993).
  %%CITATION = PHRVA,D47,1407;%%
}

\lref\fef{C. Fefferman and C.R. Graham, ``Conformal Invariants",
in {\it Elie Cartan et les Math\'{e}matiques d'aujourd'hui}
(Ast\'{e}risque, 1985) 95.}

%\BanadosWN
\lref\btz{
  M.~Banados, C.~Teitelboim and J.~Zanelli,
  ``The Black hole in three-dimensional space-time,''
  Phys.\ Rev.\ Lett.\  {\bf 69}, 1849 (1992)
  [arXiv:hep-th/9204099];  M.~Banados, M.~Henneaux, C.~Teitelboim and J.~Zanelli,
  ``Geometry of the (2+1) black hole,''
  Phys.\ Rev.\ D {\bf 48}, 1506 (1993)
  [arXiv:gr-qc/9302012].
  %%CITATION = HEP-TH 9204099;%%
}

%\BilalPH
\lref\BilalPH{
  A.~Bilal and C.~S.~Chu,
  ``A note on the chiral anomaly in the AdS/CFT correspondence and 1/N**2
  correction,''
  Nucl.\ Phys.\ B {\bf 562}, 181 (1999)
  [arXiv:hep-th/9907106];
  %%CITATION = HEP-TH 9907106;%%
  ``Testing the AdS/CFT correspondence beyond large N,''
  arXiv:hep-th/0003129.
  %%CITATION = HEP-TH 0003129;%%
}

%\MansfieldZW
\lref\MansfieldZW{
  P.~Mansfield and D.~Nolland,
  ``Order 1/N**2 test of the Maldacena conjecture: Cancellation of the
  one-loop Weyl anomaly,''
  Phys.\ Lett.\ B {\bf 495}, 435 (2000)
  [arXiv:hep-th/0005224];
  P.~Mansfield, D.~Nolland and T.~Ueno,
  ``Order 1/N**2 test of the Maldacena conjecture. II: The full bulk  one-loop
  contribution to the boundary Weyl anomaly,''
  Phys.\ Lett.\ B {\bf 565}, 207 (2003)
  [arXiv:hep-th/0208135].
  %%CITATION = HEP-TH 0208135;%%
}

%\WittenHC
\lref\WittenHC{
  E.~Witten,
  ``Five-brane effective action in M-theory,''
  J.\ Geom.\ Phys.\  {\bf 22}, 103 (1997)
  [arXiv:hep-th/9610234].
  %%CITATION = HEP-TH 9610234;%%
}

%\AharonyRZ
\lref\AharonyRZ{
  O.~Aharony, J.~Pawelczyk, S.~Theisen and S.~Yankielowicz,
  ``A note on anomalies in the AdS/CFT correspondence,''
  Phys.\ Rev.\ D {\bf 60}, 066001 (1999)
  [arXiv:hep-th/9901134].
  %%CITATION = HEP-TH 9901134;%%
}

%\BlauVZ
\lref\BlauVZ{
  M.~Blau, K.~S.~Narain and E.~Gava,
  ``On subleading contributions to the AdS/CFT trace anomaly,''
  JHEP {\bf 9909}, 018 (1999)
  [arXiv:hep-th/9904179].
  %%CITATION = HEP-TH 9904179;%%
}

%\NaculichXU
\lref\NaculichXU{
  S.~G.~Naculich, H.~J.~Schnitzer and N.~Wyllard,
  ``1/N corrections to anomalies and the AdS/CFT correspondence for
  orientifolded N = 2 orbifold models and N = 1 conifold models,''
  Int.\ J.\ Mod.\ Phys.\ A {\bf 17}, 2567 (2002)
  [arXiv:hep-th/0106020].
  %%CITATION = HEP-TH 0106020;%%
}

%\FerraraDD
\lref\FerraraDD{
  S.~Ferrara and R.~Kallosh,
  ``Supersymmetry and Attractors,''
  Phys.\ Rev.\ D {\bf 54}, 1514 (1996)
  [arXiv:hep-th/9602136].
  %%CITATION = HEP-TH 9602136;%%
}

%\NojiriMH
\lref\NojiriMH{
  S.~Nojiri and S.~D.~Odintsov,
  ``On the conformal anomaly from higher derivative gravity in AdS/CFT
  %correspondence,''
  Int.\ J.\ Mod.\ Phys.\ A {\bf 15}, 413 (2000)
  [arXiv:hep-th/9903033].
  %%CITATION = HEP-TH 9903033;%%
}

%\DeserWH
\lref\DeserWH{
  S.~Deser, R.~Jackiw and S.~Templeton,
  ``Topologically Massive Gauge Theories,''
  Annals Phys.\  {\bf 140}, 372 (1982)
  [Erratum-ibid.\  {\bf 185}, 406.1988\ APNYA,281,409
  (1988\ APNYA,281,409-449.2000)]; ``Three-Dimensional Massive Gauge Theories,''
  Phys.\ Rev.\ Lett.\  {\bf 48}, 975 (1982).
  %%CITATION = APNYA,140,372;%%
}

%\JackiwMN
\lref\JackiwMN{
  R.~Jackiw,
  ``Fifty years of Yang-Mills theory and my contribution to it,''
  arXiv:physics/0403109.
  %%CITATION = PHYS-ICS 0403109;%%
}

\lref\strings{
Talks by F. Denef, E. Verlinde, A. Strominger, and X. Yin at Strings 2006 (Beijing).}

%\KawaiJK
\lref\KawaiJK{
  T.~Kawai, Y.~Yamada and S.~K.~Yang,
  ``Elliptic Genera And N=2 Superconformal Field Theory,''
  Nucl.\ Phys.\ B {\bf 414}, 191 (1994)
  [arXiv:hep-th/9306096].
  %%CITATION = HEP-TH 9306096;%%
}

%\MooreFG
\lref\MooreFG{
  G.~W.~Moore,
  ``Les Houches lectures on strings and arithmetic,''
  arXiv:hep-th/0401049.
  %%CITATION = HEP-TH 0401049;%%
}

%\MaldacenaBP
\lref\MaldacenaBP{
  J.~M.~Maldacena, G.~W.~Moore and A.~Strominger,
  ``Counting BPS black holes in toroidal type II string theory,''
  arXiv:hep-th/9903163.
  %%CITATION = HEP-TH 9903163;%%
}

%\HansenWU
\lref\HansenWU{
  J.~Hansen and P.~Kraus,
  ``Generating charge from diffeomorphisms,''
  arXiv:hep-th/0606230.
  %%CITATION = HEP-TH 0606230;%%
}

%\ElitzurNR
\lref\ElitzurNR{
  S.~Elitzur, G.~W.~Moore, A.~Schwimmer and N.~Seiberg,
  ``Remarks On The Canonical Quantization Of The Chern-Simons-Witten Theory,''
  Nucl.\ Phys.\ B {\bf 326}, 108 (1989).
  %%CITATION = NUPHA,B326,108;%%
}

%\GukovID
\lref\GukovID{
  S.~Gukov, E.~Martinec, G.~W.~Moore and A.~Strominger,
  ``Chern-Simons gauge theory and the AdS(3)/CFT(2) correspondence,''
  arXiv:hep-th/0403225.
  %%CITATION = HEP-TH 0403225;%%
}

%%%%%%%%%%%%%%%%%%%%%%%%%%%
% Main text begins here
%%%%%%%%%%%%%%%%%%%%%%%%%%%%%%%%%%%%%%%%%%%%%%
%%%%%%%%%%%%%%%%%%%%%%%%%%%
\baselineskip14pt
\newsec{Introduction}

In situations where string theory accounts for black hole thermodynamics in
quantitative detail the microscopic theory is a conformal field theory describing
the bound states of various branes. Schematically, we  can write an
equivalence between black hole and CFT partition functions:
\eqn\ia{
Z_{\rm BH} = Z_{\rm CFT}~.
}
The early successes in establishing \ia\ involved matching the large charge
asymptotics of the two sides.
In the last few years there has been much work
(including \refs{\CardosoFP,\OSV,\Sen,\DDMP,\KrausVZ}) on refining
this identification to include the sub-leading asymptotics as well. On the black
hole side this requires the inclusion of higher order spacetime effects, such
as higher derivative corrections to the action.   In favorable cases one has
sufficient control to compute both sides of \ia\ beyond leading order  and verify that the equality
is upheld \refs{\CardosoFP,\OSV,\Sen,\DDMP,\KrausVZ}.

The identification \ia\ is most naturally viewed as an example of the AdS/CFT correspondence.
This interpretation is possible because the near horizon geometry of the black holes
considered is locally AdS$_3$~\StromBTZ. The AdS/CFT point of view explains, for
example, why the partition functions of the black hole and the CFT agree in their leading
exponential dependence.  
%This follows from the equality of the central charges, which in turn is a consequence of
%anomaly cancellation \KrausVZ.
Our working assumption is that the AdS$_3$/CFT$_2$ correspondence is responsible
also for the more detailed agreements seen when additional higher order
spacetime effects are taken into
account. Although there are ways that this assumption could fail (for example, by
AdS$_2$ playing an essential role) it seems rather conservative to us.

The leading exponential behavior of the partition function  at high (left or right moving)  temperature is determined by the
central charge.
In \KrausVZ\ we showed that central charges on the two sides must agree due
to cancellation of local anomalies. The argument is an adaptation to the AdS/CFT
correspondence of the anomaly inflow mechanism explained in \HMM\ (see also \MSW).  Since it is the
exact central charges that agree,  higher derivative corrections are taken into account.
In particular small black holes (with vanishing classical area, corresponding to vanishing
central charge at the leading order) can be considered; and the agreement also extends
to non-supersymmetric and near extremal black holes. The significance of the anomaly approach is that it is
robust, since anomalies are captured completely by a single term at one loop order.
In this way we can bypass the need to determine all the explicit higher derivative
terms in the spacetime action. 

In this paper we develop spacetime aspects of the relation \ia\ in more detail and show how
to compute the left hand side to an accuracy that extends beyond knowing the central
charge.  A key point is that Chern-Simons terms dominate the theory close to the boundary
of AdS$_3$.  This motivates
us to develop Chern-Simons theory carefully in the spirit of the AdS/CFT correspondence
and holographic renormalization.   In particular, we systematically derive the 
boundary stress
tensor and currents by studying the bulk theory in the presence of appropriate sources.   
We also consider the spacetime implementation of modular invariance and spectral
flow in detail, and we make precise statements about the lattice of currents corresponding
to particular string theory realizations.  

A major precursor to the present work is  the ``Farey tail" \farey.  These authors showed
that the elliptic genus of the D1-D5 system admits an expansion that is highly suggestive
of a supergravity interpretation in terms of a sum over geometries.   One of our main
objectives is to give a more first principles derivation of the gravitational side of this story. 
Another primary goal is to extend all this to the ${\cal N}=2$ context, which is more
sensitive to higher derivative terms and other higher order spacetime effects, and has been
the subject of much recent discussion.   We set up our formalism so that we can
consider the ${\cal N}=4$ and ${\cal N}=2$ cases in parallel.  
  The end
result is that we can give a coherent formalism for computing partition functions
from the spacetime point of view.

%but we treat the gauge fields in more detail
%and also take some detours needed to take higher derivative terms into account. The end
%result of our formalism is that we find a number of new robust results, and we give them a
%spacetime interpretation. {\bf more specific}

An important  stimulus for the recent interest in this subject was  the OSV conjecture \OSV\ that the
black hole CFT is related to the topological string through
\eqn\ib{
Z_{\rm CFT}(p^I,q_I) = \int {d\phi\over 2\pi} ~|\psi_{\rm top}(X^I = p^I + {i\over \pi}\phi^I)|^2
e^{-\pi\phi^I \cdot q_I}~.
}
Much of the work in analyzing \ib\ has taken the approach of keeping only higher derivative F-terms in a near horizon
AdS$_2\times S^2$. There is clearly a lot that is right about these conjectures but
many details remain confusing. For example, the measure of the integral
in \ib\ is unspecified and the right hand side is defined only perturbatively.
Also, since OSV interpret each of the expressions \ia-\ib\ as an index, it is not clear why
Wald's entropy formula applies. We will not resolve all these issues here but we will give
further evidence that the AdS$_3$/CFT$_2$
correspondence is a useful setting for addressing them.

We have recently learned that several related works on the OSV conjecture and the
Farey tail are in progress \refs{\strings,\GaiottoWM}. 
Although there is some overlap, our work
is complementary in that we emphasize the spacetime issues brought up
by the AdS/CFT correspondence, and the general aspects of the problem. We also
pay particular attention to situations with enhanced supersymmetry, such as
M-theory on $K3\times T^2$.

This paper is organized as follows. We begin in section 2 with a brief review of the
elliptic genus from a CFT point of view. In section 3 we introduce the ingredients
we need in our supergravity approach, with particular emphasis on gauge fields.
Section 4 contains  several simple but important examples that illustrate the
approach. In particular we reconsider the computation of the black hole
entropy in the saddle point approximation.
In section 5 we turn to our main interest, laying out the strategy for computing the
full elliptic genus from supergravity. In carrying out the computation we consider
first, in section 6, the supergravity contributions, and next, in section 7, the contributions
from wrapped branes. Finally, in section 8, we combine the various pieces and
discuss the result.

\newsec{Review: partition functions in CFT}

We begin by reviewing the definitions and properties of the CFT partition functions
that we will be trying to reproduce from supergravity/string theory.  
Our discussion here parallels that in \farey; also useful are \refs{\KawaiJK,\MooreFG}.
The focus will
be on theories with either $(4,4)$ or $(0,4)$ supersymmetry, though in fact
replacing any of the $4$'s by $2$'s makes almost no difference.    We make reference
to the two chiralities of the CFT with the convention (holomorphic, anti-holomorphic) $\sim$ (left,right). 

Besides the
Virasoro algebras, we play close attention to $U(1)$ and R-symmetry current algebras.
We write out the relevant formulas for the holomorphc currents; the anti-holomorphic
currents are included by making  the obvious substitutions.
We write $U(1)$ current algebra OPEs as
\eqn\aab{ j^I(z)j^J(0) \sim {k^{IJ} \over 2z^2}~,}
and the $SU(2)_R$ current algebra OPEs as
\eqn\aac{ J^i(z) J^j(0) \sim {k \over 2z^2}\delta^{ij} + {i\epsilon^{ijk}
\over z}J^k(0) ~. }
The leftmoving central charge of the theory is
\eqn\aaca{ c= 6k~.}

In terms of the usual expansion of the currents in modes   $j^I_n$
and $J^i_n$,  the commutation relations with the Virasoro generators $L_n$  are
\eqn\aae{\eqalign{ [L_m,L_n]& = (m-n)L_{m+n} +{c\over
12}(m^3-m)\delta_{m+n}~, \cr [L_m,j^I_{n}]&=-n j^I_{m+n}~, \cr
[j^I_{m},j^J_{n}]&=\half m k^{IJ} \delta_{m+n}~,  \cr
[L_m,J^i_{n}]&=-n J^i_{m+n}~, \cr [J^i_{m},J^j_{n}]&=\half m
k\delta_{m+n}\delta^{ij}+i\epsilon_{ijk} j^k_{m+n}~. }}

When we refer to ``charges"  we will mean
\eqn\aad{\eqalign{ J_0 &= 2 J^3_0~, \quad J_0 \in \IZ~, \cr
q^I&=2j^I_0~,}}
in a basis that diagonalizes the operators.

The combined Virasoro/current algebras admit the following
spectral flow automorphisms:
\eqn\ad{\eqalign{L_0 &~\rightarrow ~L_0 + \eta J_0 + k\eta^2~, \cr
J_0 &~\rightarrow~ J_0 + 2k \eta~, }}
in the $SU(2)$ case, and
\eqn\hq{ \eqalign{L_0 &~\rightarrow ~L_0 + \eta_I q^I +
k^{IJ}\eta_I \eta_J~, \cr q^I &~\rightarrow~ q^I + 2 k^{IJ} \eta_J~,
}}
in the $U(1)$ case.    Integer $\eta$ preserves fermion periodicities,
while half-integer $\eta$ interchanges the NS and R sectors.
There is an analogous statement for the $\eta^I$, depending on the
relevant charge lattice, $q^I \in \Gamma$.

\subsec{Elliptic genus}

To define the elliptic genus we introduce potentials for the charges $J_0$ and
$q^I$.   Since these charges appear symmetrically, it is natural to relabel the
R-charge as
\eqn\hqa{ q^0 \equiv J_0~,}
so that $I=0,1,2,\ldots$.  We also extend the definition of $k^{IJ}$ such that
$k^{00} = k$, $k^{0,I>0}=k^{I>0,0}=0$.

The elliptic genus is now defined as\foot{In the $(0,4)$ case $RR$ is replaced by $R$.}
\eqn\aa{\eqalign{ \chi(\tau,z_I)&= \Tr_{RR} \left[ e^{2\pi i \tau
(L_0 -c/24)} e^{-2\pi i \taub (\Lt_0-\tilde{c}/24)} e^{2\pi i z_I q^I}(-1)^F\right]
% \cr &= \sum_{n\geq 0, r, r^I} c(n,r,r^I) q^n y^r y_I^{r^I}
~.}}
 Only rightmoving ground states, with
$\Lt_0-\tilde{c}/24=0$, contribute, so the elliptic genus
does not depend explicitly on $\taub$.    On the other hand, all
leftmoving states can contribute.   The elliptic genus is invariant under
smooth deformations of the CFT.  This follow from the quantization of the
charges and of $L_0 - \Lt_0$, together with the fact that only rightmoving
ground states contribute.

In certain cases relevant for AdS/CFT the index defined in \aa\ vanishes,
and one needs to consider a modified index by inserting a factor of $\tilde{F}^2$ \MaldacenaBP. 
An example is for the case of the $(0,4)$ CFT of wrapped M5-branes, where
the vanishing comes from the contribution of the center of mass multiplet.
As we'll discuss later, the center of mass degrees of freedom are absent in
the AdS description, and so from the bulk point of view we can compute
the index as defined in \aa\ and obtain a nonvanishing result.   What we will not
discuss here is the precise relation of this bulk index to the modified CFT index,
although this clearly deserves further study.  

We now state the main general properties of the elliptic
genus.

\vskip.2cm \noindent {\bf Modular transformation}
\eqn\ab{ \chi({a\tau + b \over c\tau +d},{z_I
\over c\tau +d}) =e^{2\pi i  {c z^2 \over c\tau +d} } \chi(\tau,z_I)~,}
where we define
\eqn\abz{ z^2 \equiv k^{IJ} z_I z_J~.}
Note that $c$ in \ab\ is {\it not} the central charge!

In Appendix A we review the proof of \ab.  It is most easily understood in terms
of the relation between the Hamiltonian and path integral expression for the
elliptic genus.   There is  a natural  modular invariant path integral expression,
but the Hamiltonian corresponding to this action differs from that appearing in
the exponential of \aa.   The modular transformation of this extra factor yields
the prefactor in \ab.

\vskip.2cm \noindent {\bf Spectral flow}

The spectral flow automorphisms imply the relation
\eqn\ac{\chi(\tau,z_I+\ell_I \tau+m_I)= e^{-{2\pi i} ( \ell^2\tau  +2 \ell \cdot z)}\chi(\tau,z_I)~,
  }
where $m_I$ obeys $m_I q^I \in \IZ$, and we defined $\ell^2 = k^{IJ}\ell_I \ell_J$, $\ell \cdot z = k^{IJ} \ell_I z_J$.

This also implies that if we expand the elliptic genus as
\eqn\acb{\chi(\tau,z_I) = \sum_{n, r^I} c(n,r^I)
e^{2\pi i n \tau + 2\pi i z_I r^I}~, }
then the expansion coefficients are a function of a
single spectral flow invariant combination:
\eqn\ada{ c(n,r^I) = c(n-{r^2 \over 4}) ~.}
Here we defined $r^2 = k_{IJ} r^I r^J$, where $k_{IJ}$ denotes the inverse of
$k^{IJ}$.

\vskip.2cm \noindent {\bf Factorization of dependence on
potentials}

We can explicitly write the dependence of the elliptic genus on
the potentials $z_I$.  The intuition behind this is that we
can always separate the CFT into the currents  plus everything
else, and the current part can
be realized in terms of free bosons.  We have:
\eqn\adz{\chi(\tau,z_I) = \sum_{\mu^I}
h_{\mu}(\tau) \Theta_{\mu,k}(\tau,z_I)~,}
with
\eqn\aez{\eqalign{  \Theta_{\mu,k}(\tau,z_I)&=   \sum_{\eta_I}
e^{ {i\pi \tau \over 2}(\mu+2k\eta )^2}  e^{2\pi i z_I(\mu^I+2k^{IJ}\eta_J)}~.}}
We are using the shorthand notation
\eqn\aeza{  (\mu+2k\eta )^2 \equiv k_{IJ}(\mu^I + 2k^{IK}\eta_K)(\mu^J + 2 k^{JL}\eta_L)~.}

The combined sum over $\mu^I$ and $\eta_I$ includes the complete spectrum
of charges.  The sum over $\eta_I$ corresponds to shifts of the charges by
spectral flow, and so the sum on $\mu_I$ is over a fundamental domain with respect
to these shifts.

\vskip.2cm \noindent {\bf NS sector elliptic genus}

Using spectral flow, we can alternatively write the elliptic genus in the NS
sector.    In particular, by performing a half-integer spectral flow on the R-symmetry
charges, we obtain
\eqn\adb{\eqalign{ \chi_{RR}(\tau,z_0,z_I) &=e^{-2\pi i kz_0}
\chi_{NS,NS}(\tau,z_0-\half \tau,z_I)~, \cr
\chi_{NS,NS}(\tau,z_0,z_I)&=\Tr_{NS,NS}\left[e^{2\pi i\tau L_0}e^{
-2\pi i \taub(\Lt_0-\half \Jt_0)}e^{2\pi i (z_0 q^0+ z_I q^I)}(-1)^F
\right]~,}}
where in the above $I=1,2,\ldots$.
$\chi_{NS,NS}$ receives contributions from chiral primaries, $\Lt_0-\half \Jt_0=0$.
These chiral primaries are, as usual, the spectral flows of R ground states.

\vskip.2cm \noindent {\bf Farey tail expansion}

The main observation
of \farey\  was that  upon applying the ``Farey tail transform", the elliptic genus
admits an expansion that is suggestive of a supergravity interpretation in terms
of  a sum over geometries.   We will essentially state the result here, referring
to \farey\ for the detailed derivation.

The properties \ab\ and \ac\ are the definitions of a ``weak Jacobi form" of
weight $w=0$ and index $k$.  Actually, the definition strictly applies when
$k$ is a single number rather than a matrix, but we will still use this langauge.

The Farey tail transformed elliptic genus is
\eqn\mb{ \chit(\tau,z_I)= \left({1 \over 2\pi i}\p_\tau-{1 \over 4}
{\p_z^2\over (2\pi i)^2}\right)^{3/2}\chi(\tau,z_I)~,}
where $\p_z^2 = k_{IJ} \p_{z_I} \p_{z_J}$.  $\chit$
is a weak Jacobi form of weight $3$ and index $k$, and admits the expansion
\eqn\mc{\chit(\tau,z_I)=e^{-{\pi z^2 \over \tau_2}}\sum_{\Gamma_\infty\setminus\Gamma}{1\over
(c\tau+d)^3 }\hat{\chi}\left( {a\tau+b\over
c\tau+d},{z_I\over ct +d}\right)~, }
with
\eqn\md{\eqalign{ \hat{\chi}(\tau,z_I) &=e^{{\pi z^2 \over \tau_2}}
\hat{\sum_{\mu,\tilde{\mu},m,\tilde{m}}}
\tilde{c}(m,\mu^I)e^{2\pi i (m-{1 \over
4}\mu^2)\tau}\Theta_{\mu,k}(\tau,z_I)~,}}
and $\Theta_{\mu,k}(\tau,z_I)$ was defined in  \aez.   The hatted summation appearing
in \md\ is over states with $m-{1 \over 4}\mu^2 <0$. From the gravitational point
of view these will be states below the black hole threshold and the sum over $\Gamma_\infty\setminus\Gamma$ then adds the black holes back in. In mathematical
terminology \md\ defines $\hat{\chi}$ as the ``polar part" of the elliptic genus.
   The coefficients $\tilde{c}(m,\mu^I)$ in
\md\ are related  to those in \acb\ by
\eqn\mda{ \tilde{c}(m,\mu^I) = (m-{\mu^2 \over 4})^{3/2} c(m-{\mu^2 \over 4})~,}
as follows from \mb\ and from using \ada.  The main point is that the transformed
elliptic genus $\tilde{\chi}$ can be reconstructed in terms of its polar part
$\hat{\chi}$.

\subsec{General partition function}

If we include potentials for both left and  right moving charges  we define a partition function that
receives contributions from all states of both chiralities.    This object is no longer
invariant under deformations of the CFT, and we have little hope of computing it
exactly.  Nevertheless, we can infer some general properties, and  do an
approximate computation in the regime of weakly coupled supergravity.

We define
\eqn\me{  Z(\tau,z_I; \taub,\zt_I)= \Tr_{RR} \left[ e^{2\pi i \tau
(L_0 -c/24)} e^{-2\pi i \taub (\Lt_0-\tilde{c}/24)} e^{2\pi i z_I q^I}e^{-2\pi i \zt_I \qt^I}\right]
~.}
The properties of the elliptic genus that we reviewed above all have their obvious
analogs for the partition function \me.

\newsec{Supergravity analysis: preliminaries}

Having reviewed the definitions and properties of the CFT partition functions,
we now turn to their study in supergravity.  The CFTs in question are dual to
string theories on AdS$_3$ times some compact space.   In general, these theories
have all manner of complications from higher derivative terms and massive
string/brane states.   Fortunately, for computing the elliptic genus not all of this information
is necessary, and we can get remarkably far by taking advantage of all the symmetries,
and by carefully studying the long-distance part of the theory.  The remaining input
which will still be needed  to complete the calculation is the spectrum of massive
string/brane states, which can be computed exactly in certain cases.  We 
proceed step-by-step, upward in energy scales.  We start with the low
energy effective action where the relevant {\it massless} fields on AdS$_3$ are the 
metric and a collection of gauge fields. Later we discuss supergravity Kaluza-Klein 
modes and finally nonperturbative states.  

\subsec{Gravity action}

The action for the metric is
%\foot{In three dimensions the gravitational action can
%famously be rewritten in the form of a Chern-Simons theory with non-compact
%gauge group. This has the advantage that gravity and gauge fields are treated
%on the same footing. Here we stick to the more conventional formulation of gravity
%because various subtleties are better understood there. The two formalisms
%are entirely equivalent.}
%
\eqn\aff{ S_{grav} = {1\over 16\pi G}\int\! d^3x \sqrt{g}(R-{2 \over
\ell^2}) + \int\! d^3x~ \Omega_3(\Gamma) +{1 \over 8\pi G}\int_{\p AdS}\! d^2x \sqrt{g} \left( \Tr K -{1 \over \ell}\right)+ \ldots~. }
We work in Euclidean signature.   The second term is a gravitational Chern-Simons term,
$\Omega_3(\Gamma) = \beta \Tr( \Gamma d\Gamma +{2 \over 3} \Gamma^3)$.
It plays a crucial  role in theories with $c \neq \tilde{c}$ (specifically,
$c-\tilde{c} = 96 \pi \beta$) and its effects were studied in \KrausVZ.
The next terms are the Gibbons-Hawking and
boundary counterterms that are standard for gravity in asymptotically AdS$_3$
spacetimes.    Also indicated by the $\ldots$ are the higher derivative terms that are present in string theory constructions. Although these certainly contribute, even without knowing
their precise form we can incorporate all their effects provided we
carefully implement all symmetries and anomalies.

The statement that the metric is asymptotically AdS$_3$ means that it takes
the Fefferman-Graham form
\eqn\afg{ ds^2 = d\eta^2 +e^{2\eta/\ell} g^{(0)}_{\alpha\beta}
dx^\alpha dx^\beta + g^{(2)}_{\alpha\beta} dx^\alpha dx^\beta +
\ldots~.}
Here $g^{(0)}_{\alpha\beta}$ is the ``conformal boundary metric".    The boundary stress
tensor is defined by computing the on-shell variation of the action with respect to the
conformal boundary metric \balkraus
\eqn\ae{ \delta S = {1 \over 2} \int_{\p AdS}\! d^2x \sqrt{g^{(0)}}T^{\alpha\beta} \delta g^{(0)}_{\alpha\beta} ~,}
yielding
\eqn\aea{ T^{grav} _{\alpha\beta} = {1 \over 8\pi G\ell} \left(
g^{(2)}_{\alpha\beta}- \Tr (g^{(2)}) g_{\alpha\beta}^{(0)}\right) +{\rm( higher~deriv.)}~.}
We added the {\it grav} superscript because the stress tensor also receives a
contribution from the gauge fields discussed in the following.

\subsec{Gauge field action}

\vskip.2cm \noindent {\bf Leftmoving $SU(2)$ currents}

Next, consider the gauge fields.   Associated with $4$ left or right moving supercharges
is an $SU(2)$ current algebra that is realized in the bulk by $SU(2)$ gauge fields.
The $SU(2)$ gauge fields have a Chern-Simons term as the leading long-distance part of
the action, with a coefficient related to the level of the current algebra. 
See \refs{\ElitzurNR,\GukovID } for some relevant earlier work on Chern-Simons theory.  
 For the leftmoving
gauge fields we write
\eqn\afh{ S_{gauge} =-{ik\over 4\pi} \int \! d^3x \,\Tr(AdA+{2\over 3} A^3) + S^{bndy}_{gauge}~,}
with $A = A^a {i\sigma^a \over 2}$.  From a higher dimensional point of view,
 the $SU(2)$ gauge fields can be thought of 
 as being associated with the isometries of
a sphere, and the Chern-Simons term \afh\ can be derived by dimensional reductiion
\HansenWU.   Invariance of the path integral under
large gauge transformation fixes $k$ to be an integer, and we'll rederive below the
standard fact that $k$ is the level of the boundary current algebra.  Supersymmetry
relates $k$ to the leftmoving central charge as $c=6k$.    The boundary term indicated
in \afh\ and given explicitly below is necessary in order that the currents have only leftmoving components.

The gauge fields admit the expansion
\eqn\afi{ A = A^{(0)} + e^{-2\eta/\ell} A^{(2)} +\ldots~,}
and we choose the gauge $A_\eta =0$.    Analysis of the field
equations (including the effect of Maxwell type terms) shows  that
$A^{(0)}$ is a flat connection; that is, the field strength
corresponding to \afi\ falls off as $e^{-2\eta/\ell}$.    The boundary current
is obtained from the on-shell variation of the action with respect to $A^{(0)}$
\eqn\afj{ \delta S = {i\over
2\pi}  \int_{\p AdS}\! d^2x \sqrt{g^{(0)}}J^{\alpha a} \delta A^{(0)a}_\alpha ~.}
We expect the boundary current corresponding to $SU(2)_L$ to be purely
leftmoving. In a general coordinate system this amounts to the imaginary
anti-self dual condition $\star J = -iJ$, with $\star$ defined with respect to
$g^{(0)}_{\alpha\beta}$. (The opposite sign holds for the rightmoving
case). However, the variation of the action \afh\ does not give a purely
leftmoving current unless we take the boundary term as
\eqn\afk{ S_{gauge}^{bndy} =- {k \over 16\pi} \int_{\p AdS} \!
d^2 x \sqrt{g} g^{\alpha\beta} A^{a}_\alpha A^{a}_\beta~.}
This yields the imaginary anti-self dual current
\eqn\afl{\eqalign{J^a_\alpha & =
{ik\over 4} (A^{(0) a}_\alpha -i \epsilon_\alpha^{~\beta}
A^{(0)a}_\beta )~.
 }}
In conformal gauge,  $g^{(0)}_{\alpha\beta}dx^\alpha dx^\beta = dw d\wb$, \
we find
\eqn\afm{ J^a_w = {ik \over 2} A^{(0)a}_w~,\quad J^a_{\wb} =0~.}
This is an exact expression for the current, uncorrected by the higher derivative
terms in \afh, because of the flatness of $A^{(0)}$.   This will be important.

An equivalent way to motivate the boundary term \afk\ is to demand that in the
variational principle we only fix boundary conditions for $A^{(0)a}_{\wb}$ and not
$A^{(0)a}_{w}$.   Fixing both components is too strong in that there will typically
not be any smooth solutions in the bulk with the chosen boundary conditions.
The condition that the variation of the action takes the form 
$\delta S \sim \int J_w \delta A^{(0)}_{\wb}$
then leads to the same conclusions as above.

\vskip.2cm \noindent {\bf Rightmoving $SU(2)$ currents}

The rightmoving gauge fields are described by the action
\eqn\afn{ S_{gauge} ={i\kt \over 4\pi} \int \! d^3x \,\Tr(\At d\At+{2\over 3} \At^3) -
{\kt \over 16\pi} \int_{\p AdS} \!
d^2 x \sqrt{g} g^{\alpha\beta} \At^{a}_\alpha \At^{a}_\beta~.}
Note that the Chern-Simons term appears with an opposite sign from \afh, and the
boundary term was fixed by demanding that the current be purely rightmoving:
\eqn\afo{\eqalign{\Jt^a_\alpha & =
{i\kt\over 4} (\At^{(0) a}_\alpha +i \epsilon_\alpha^{~\beta}
\At^{(0)a}_\beta )~,
 }}
or
\eqn\afp{ \Jt^a_w =0~,\quad J^a_{\wb} = {i\kt \over 2} \At^{(0)a}_{\wb}~.}

\vskip.2cm \noindent {\bf Gauge field contribution to stress tensor}

The gauge field boundary terms  are metric dependent and hence contribute
to the stress tensor as:\foot{The variation of the metric is computed while holding
fixed the lower components of the gauge fields.}
\eqn\afq{ T^{gauge}_{\alpha\beta}  = {k \over 8\pi}(A^{(0)a}_\alpha A^{(0)a}_\beta -\half A^{(0)a
\gamma}A^{(0)a}_\gamma g^{(0)}_{\alpha\beta}) ~+~ \left[(k,A) \rightarrow (\kt,\At)\right]~,}
or
\eqn\afr{\eqalign{ T^{gauge}_{ww} &= {k \over 8\pi} A^{(0)a}_w A^{(0)a}_{w} +
 {\kt \over 8\pi} \At^{(0)a}_w \At^{(0)a}_{w}~, \cr
T^{gauge}_{\wb \wb} &= {k \over 8\pi} A^{(0)a}_{\wb} A^{(0)a}_{\wb} + {\kt \over 8\pi} \At^{(0)a}_{\wb} \At^{(0)a}_{\wb}~,
\cr
 T^{gauge}_{w \wb}&=T^{gauge}_{ \wb w}=0~.}}
The index (0) on the gauge field reminds us that boundary expressions strictly refer to 
just the leading term in the Fefferman-Graham  expansion \afi\ for the bulk gauge field.
In the following we will reduce clutter by dropping this index.

 \vskip.2cm \noindent {\bf $U(1)$ currents}

 Besides the $SU(2)$ currents, we will typically have some number of left and right
 moving $U(1)$ currents. Only those gauge fields that appear in Chern-Simons terms will
 contribute to boundary currents.
     The Chern-Simons term has the form $S \sim \int C^{IJ}A_I dA_J$.
 By a change of basis we can put $C^{IJ}$ in block diagonal form,
 \eqn\afs{ C^{IJ}=\left( \matrix{ k^{IJ} & 0 \cr 0& \kt^{IJ}}\right)~,}
 where $k^{IJ}$ ($\kt^{IJ}$) has positive(negative) eigenvalues.   We then write the
relevant part of the action as
\eqn\aft{S=  {i \over 8\pi} \int \! d^3x \, (k^{IJ} A_I dA_J-\kt^{IJ} \At_I d\At_J) -
{1 \over 16\pi} \int_{\p AdS} \!
d^2 x \sqrt{g} g^{\alpha\beta} (k^{IJ}A_{I
\alpha} A_{J\beta}+\kt^{IJ}\At_{I\alpha}
\At_{J\beta} )   ~.}

We combine this with the $SU(2)$ gauge fields as follows.   We will be considering
solutions in which in which only the $a=3$ component of $A^{(0)a}$ and $\At^{(0)a}$
is nonvanishing.   Then, so far as getting the correct currents is concerned, these can
thought of as $U(1)$ currents, and can be incorporated in \aft\ by extending the $I$
indices to include $I=0$ with
\eqn\afu{  A^{(3)} = A_{I=0}~,\quad \At^{(3)} = \At_{I=0}~.}
With this in mind, we can now take \aft\ as our general gauge field action.    We will
write it in condensed form as
\eqn\afv{S=  {i \over 8\pi} \int \! d^3x \, (AdA-\At d\At) -
{1 \over 16\pi} \int_{\p AdS} \!
d^2 x \sqrt{g} g^{\alpha\beta} (A_\alpha A_\beta+\At_\alpha
\At_\beta )   ~,}
where the appropriate index contractions with $k^{IJ}$ and $\kt^{IJ}$ are implicit.

In conformal gauge, the gauge fields contribute to the currents and stress tensor as,
\eqn\afw{\eqalign{ T^{gauge}_{ww} &= {1 \over 8\pi} A_{w}^2+
 {1 \over 8\pi} \At^2_{w} ~, \cr
T^{gauge}_{\wb \wb} &=  {1 \over 8\pi} A^2_{\wb}+
{1 \over 8\pi} \At^2_{\wb}~,
\cr
 T^{gauge}_{w \wb}&=T^{gauge}_{ \wb w}=0~, \cr
 J^{I}_{ w} &= {i\over 2}k^{IJ} A_{Jw}~,\quad
 J^{I}_{\wb}=0~, \cr \Jt^{I}_{ w}&=0~,\quad  \Jt^{I}_{ \wb} = {i\over 2}\kt^{IJ} \At_{J \wb}
 ~.}}
 Again, the appropriate contraction of indices is implicit.

\subsec{Anomalies}

The currents defined in \afw\ satisfy
\eqn\afx{\eqalign{  \p_{\wb} J^{I}_{w}&= {i \over 2}k^{IJ} \p_w A_{J \wb}~,\cr
\p_{w} \Jt^{I}_{\wb}&= {i \over 2}\kt^{IJ} \p_{\wb} \At_{J w}~,}}
where we used flatness of the boundary potentials.    By comparing with the chiral
anomalies of the boundary CFT, we see that $k_{IJ}$ can be identified with the $k_{IJ}$ matrix
appearing in Section 2 (and similarly for $\kt_{IJ}$).   Since they are related to chiral
anomalies, for a given string/brane realization of the AdS$_3$ geometry we can compute
$k_{IJ}$ and $\kt_{IJ}$ exactly.   By the anomaly inflow mechanism, one further knows
that they must agree with their CFT counterparts, otherwise there will be an inconsistency
in coupling the string/brane system to bulk fields.   We refer to \KrausVZ\ for the detailed
story.

\subsec{Charges}

Charges are defined as contour integrals around the AdS$_3$ boundary cylinder.
We work with the complex boundary coordinate $w\cong w+2\pi$.     The $U(1)$
charges are then
\eqn\afy{ \eqalign{q^I &= 2\oint {dw \over 2\pi i} J^I_w = i \oint
{dw \over 2\pi i} k^{IJ} A_{Jw}~,\cr \qt^I &=-2 \oint {d\wb \over 2\pi i} \Jt^I_{\wb} = -{i}  \oint
{d\wb \over 2\pi i} \kt^{IJ} \At_{J\wb}~,}}
where the factors of $2$ were included to agree with our convention in \aad.
Similarly, the Virasoro zero mode generators are\foot{Note that our normalization of the
stress tensor differs by a factor $-2\pi$ from the standard CFT definition in, e.g.,
\PolchinskiRQ.}
\eqn\afz{ \eqalign{L_0^{gauge} &= \oint dw ~T^{gauge}_{ww}=
{1 \over 8\pi}\oint dw~ A^2_{w}~,   \cr 
\Lt_0^{gauge} &=\oint d\wb~T^{gauge}_{\wb \wb}
=  {1 \over 8\pi} \oint d\wb~ \At^2_{w}
  ~.}}

More generally, we can define all the modes of the currents, $J^I_n$, and stress tensor, $L_n$,
 by inserting
factors of $w^n$ into the above integrals.   These modes then satisfy the commutation relations
in \aae.

\subsec{Spectral flow}

Spectral flow corresponds to a constant shift in the gauge potentials.  This is equivalent
to shifting the periodicities of charged fields.
%Let us adjust the normalization of  $k^{IJ}$ so that $q^I \in \IZ$.
  In the presence of a nonzero potential, the holonomy
associated with a charged particle taken around the AdS$_3$ boundary cylinder is
\eqn\aga{ e^{\half i q^I \oint dw \, A_{Iw }}~,}
so that the shift,
\eqn\agb{ A_{Iw} \rightarrow  A_{Iw} + 2\eta_I~, }
introduces the phase factor $e^{2\pi i q^I \eta_I}$. The  factor of ${1\over 2}$
in the exponent of \aga\ came from our definition of charge \afy.

Under the shift \agb\ we have,
\eqn\hqa{ \eqalign{L_0 &~\rightarrow ~L_0 + \eta_I q^I +
k^{IJ}\eta_I \eta_J~, \cr q^I &~\rightarrow~ q^I + 2 k^{IJ} \eta_J~,
}}
in agreement with \hq.  We also have the analogous formulas for a rightmoving spectral flow.

While we can perform a spectral flow with respect to any of the $U(1)$ currents,
the terminology is often reserved for the R-symmetry.  Recalling that the R-symmetry charge
is $q^0$, such a spectral flow correspond to $\eta_0$.      We pass back and forth between
the NS and R sectors by flipping the periodicity of the supercurrent; this corresponds to
taking $\eta_0 = m+\half$, with $m\in \IZ$.

\newsec{Supergravity analysis: explicit examples}

Before considering the computation of  partition functions we 
illustrate the above results with some simple examples.

\subsec{NS-NS vacuum}

The NS-NS (or simply NS in the case of $(0,4)$ susy) vacuum is invariant
under $SL(2,\IR)\times  SL(2,\IR)$.  In other words, it is invariant under the full
group of AdS$_3$ isometries, which means that it is precisely global AdS$_3$,
\eqn\agc{ ds^2 = (1 +r^2/\ell^2)\ell^2dt^2 +{dr^2 \over 1+r^2/\ell^2} +r^2 d\phi^2~.}
The contractibility of the $\phi$ circle forces the fermions to be anti-periodic in $\phi$.
Invariance under the isometry group means that this geometry has
\eqn\agd{ L_0 = \Lt_0 =0~.}

\subsec{Spectral flow to the R sector}

As described at the end of section 3.5,
to get R sector geometries we take the geometry \agc\ with
\eqn\age{ A_{0w} = 1~,}
and fermions to be periodic in $\phi$. The gauge field contribution \afr\ increases
the stress tensor from \agd\ to
\eqn\agf{L_0 = {k\over 4} = {c\over 24}~.}
Since the charge \afy\ is
\eqn\agfa{
q^0 = k ={c\over 6}~,}
this is the maximally charged R vacuum state.  To get the  maximally negatively
charged R vacuum we flip the sign in \age.    The rightmoving side is treated
analogously.

\subsec{Conical defects}

A more general  class of RR vacua are the conical defect
geometries \refs{\MaldacenaDR,\BalasubramanianRT}.
For these we take
\eqn\agg{ \eqalign{ ds^2 &= ({1 \over N^2}+{r^2 \over \ell^2})\ell^2dt^2 +{dr^2
\over
 ({1 \over N^2}+{r^2 \over \ell^2})} +r^2 d\phi^2 ~, \cr A_{0w} &= \At_{0\wb} ={1 \over N}~,}}
with $N\in \IZ$.   The angular coordinate $\phi$ has the standard $2\pi$ periodicity,
and fermions are taken to be periodic in $\phi$.

To read off the Virasoro charges we just note that by rescaling coordinates all
these geometries are locally equivalent to the $N=1$ case discussed above
in section 4.2.   In the $N=1$ case the stress tensor vanishes, and it will clearly
continue to vanish after rescaling coordinates. Thus
\agf\ still applies and so $L_0 = {k\over4}$ and $\Lt_0 = {\kt \over 4}$ as before.
The R-charge is read off from \afy\
\eqn\agh{ q^0 = {k\over N}~,\quad \qt^0 = {\kt \over N}~.}
Upper and lower bounds on $N$ are given by the quantization of R-charge.
For example, in the D1-D5 case, the condition that $q^0, \tilde{q}^0$ are integral
gives $|N|\leq N_1 N_5$ since $k=\kt=N_1 N_5$.

These conical defect geometries are singular at the origin unless the holonomy is
$\pm 1$, which corresponds to $N=\pm 1$.  In the context of the D1-D5 system,
 the singular geometries  are known to be  physical in that the singularity corresponds to the presence of $N$ coincident  Kaluza-Klein monopoles.    Another way of viewing this is that these singular
 geometries are special limits of the much larger class of smooth RR vacua geometries
 that have been heavily studied in recent years \refs{\LuninBJ,\MathurZP}.

We also note that any of the RR-vacua in \agg\ can be spectral flowed to the NSNS
sector to give chiral primary geometries.

\subsec{Black holes}

We now consider black hole geometries, and give a simple derivation of the entropy
of charged black holes that incorporates higher derivative corrections.   This is
a slight refinement of our earlier work \KrausVZ.    We use the well known trick of
relating black holes to thermal AdS by a modular transformation; the main novelty
here is the inclusion of charge and higher derivative corrections.

The starting point is global AdS$_3$, as in \agc. The complex boundary
coordinate is $w= \phi +i t/\ell$, and we identify $w\cong w+ 2\pi \cong w+2\pi \tau$.
To add charge we also want to turn on flat potentials for the gauge fields.   Now, 
the $\phi$ circle is contractible in the bulk, so to avoid a singularity at the origin we 
need to set to zero the $\phi$ component of all potentials.  We therefore allow
nonzero $A_{Iw} = - A_{I\wb}$,  and $\At_{I\wb} = - \At_{Iw}$.  
%Since we
%are considering black holes in the NS sector, fermions are anti-periodic in $\phi$.

What is the action associated with this solution? From the discussion in section 3
we know the exact expressions for the stress tensor and currents
\eqn\ahc{\eqalign{ T_{ww}&=-{k \over 8\pi} +{1 \over 8\pi} A_w^2+{1 \over 8\pi} \At_w^2~,  \cr
T_{\wb\wb}&= -{\kt \over 8\pi} +{1 \over 8\pi}A_{\wb}^2+{1 \over 8\pi}\At_{\wb}^2~, \cr
J^I_w &= {i\over 2} k^{IJ} A_{Jw}~, \cr \Jt^I_{\wb}& = {i\over 2} \kt^{IJ} \At_{J\wb}~.}}
To obtain the exact action from these formulae we need to integrate the equation
\eqn\aha{ \delta S = \int_{\p AdS}\! d^2x \sqrt{g^{(0)}}\left( {1 \over 2}T^{\alpha\beta} \delta g^{(0)}_{\alpha\beta}+{i\over 2\pi} J^{I\alpha} \delta A_{I\alpha} \right)~.}

In doing so, let us first note that the conformal gauge used hitherto fixed the conformal 
boundary metric as $dw d\wb$ and encoded the conformal structure in the periodicities 
of the coordinates. To exploit \aha\ it is advantageous to define a new coordinate $z$
\eqn\ahb{ z= {i-\taub \over \tau -\taub}w - {i-\tau\over \tau -\taub}\wb~,}
with fixed periodicity $z \cong z+2\pi \cong z+ 2\pi i$. In terms of
this coordinate the conformal structure is encoded in the metric 
\eqn\ahca{
ds^2 = dw d\bar{w} = \left| {1-i\tau\over 2} dz + {1+i\tau\over 2} d{\bar z}\right|^2~.
}
Using the new coordinates to compute the variations $\delta g^{(0)}_{\alpha\beta}$ 
with respect to $\tau$ and $\bar{\tau}$, and also transforming the $z$ and ${\bar z}$ 
components of $T^{\alpha\beta}$ back to the original coordinates, we 
can rewrite \aha\ as
\eqn\ahcb{
\delta S = (2\pi)^2 i \left[ -T_{ww} \delta\tau + T_{{\bar w}{\bar w}}\delta {\bar\tau}
+ {\tau_2\over\pi} J^I_w \delta A_{I\wb} + {\tau_2\over\pi} \Jt^I_{\wb} \delta \At_{Iw} 
\right]_{\rm const}~.
}
The {\it const} subscript  indicates that we keep just the zero mode part. 
Inserting \ahc\ into this equation we can now integrate and find our desired action as
\eqn\ahcc{ S= {i \pi k \over 2}\tau -{i\pi \kt \over 2} \taub
+\pi \tau_2 (A_{\wb}^2 + \At_w^2)~.}

A simpler derivation of this result is to just compute \ahc\ by directly evaluating the action on the 
solution. The gauge field contribution just comes from the boundary term \afk. The reason 
we proceeded in terms of \aha\ was to emphasize that the result \ahc\ is exact for an arbitrary 
higher derivative action, and also because we will generalize this computation later. 

The result \ahc\ is the action for the AdS$_3$ ground state with a flat connection
turned on. Next, we perform the modular transformation $\tau \rightarrow -1/ \tau$
in order to reinterpret the solution as a Euclidean black hole. This is implemented 
by
\eqn\ahd{ w \rightarrow -w/\tau, \quad A_{I\wb} \rightarrow -{\bar\tau} A_{I\wb}~,\quad \At_{Iw} \rightarrow - \tau \At_{Iw}~.}
The action is of course invariant since we are just rewriting it in new variables.
Using $\taub/\tau = 1- 2i \tau_2/\tau$ and introducing the potentials 
$z = -i\tau_2 A_{\wb}$ and $\zt = i\tau_2 \At_w$ (defined in 
equation A.7) we can present the result as
\eqn\ahe{\eqalign{ S&= -{i\pi k \over 2 \tau}+{i\pi \kt \over 2\taub}
-{2\pi i\tau_2^2 A_{\wb}^2 \over \tau} + {2\pi i\tau_2^2 \At_{\wb}^2 \over \taub}+\pi \tau_2 (A_{\wb}^2 +\At_w^2)
\cr & =-{i\pi k \over 2 \tau}+{i\pi \kt \over 2\taub} +
{2\pi i z^2 \over \tau} - {2\pi i \zt^2 \over \taub}-{\pi\over \tau_2}   (z^2+\zt^2)  ~.}}
This is the Euclidean action of a black hole with modular parameter $\tau$ and
potentials specified by $z$ and $\zt$.

Our result \ahe\ is the leading saddle point contribution to the path integral. As we discuss 
in appendix A, the standard canonical form of the partition function, 
defined as a trace, is related to the path integral as 
\eqn\ahf{\eqalign{ Z &=e^{-{ \pi \over \tau_2} (z^2 +\zt^2)}  Z_{PI} =  
e^{-{ \pi \over \tau_2} (z^2 +\zt^2)}\sum e^{-S}  ~.}}
The exponential prefactor cancels the last term in \ahe\ so that 
\eqn\ahg{ \ln Z =  {i\pi k \over 2 \tau}-{i\pi \kt \over 2\taub} -
{2\pi i z^2 \over \tau} + {2\pi i \zt^2 \over \taub}~,}
on the saddle point. We define the entropy $s$ by writing the partition 
function as
\eqn\agh{ Z = e^{s}  e^{2\pi i \tau
(L_0 -c/24)} e^{-2\pi i \taub (\Lt_0-\tilde{c}/24)} e^{2\pi i z_I q^I} e^{-2\pi i \zt_I q^I}~,}
where we assume that $Z$ is dominated by a single a single charge configuration 
with, e.g.,  $q^I = {1 \over 2\pi i } {\p \over \p z_I  } \ln Z$. 

Putting everything together we read off the black hole entropy as
\eqn\agi{ s= 2\pi \sqrt{{c \over 6} (L_0 - {c \over 24} -{1 \over 4}q^2)  }
+ 2\pi \sqrt{{\ct \over 6} (\Lt_0 - {\ct \over 24} -{1 \over 4}\qt^2)  }~. }
The expression \agi\ gives the entropy for a general nonextremal, rotating, charged,
black hole in AdS$_3$, including the  effect of higher derivative corrections as
incorporated in the central charges. Since we used the saddle point approximation
the formula is only valid to leading order in $\Lt_0 - {\ct \over 24} -{1 \over 4}\qt^2$;
including the subleading contribution is the topic of the next section.  It is striking 
that we have control over higher 
derivative corrections to the entropy even for nonsupersymmetric black holes. In 
earlier work \KrausVZ\ we explained this in terms of anomalies, and showed that \agi\ is in 
precise agreement with the microscopic entropy counting coming from brane 
constructions.

\newsec{Computation of partition functions in supergravity}

Let's now look at the supergravity computation of the elliptic genus
\eqn\ia{ \chi(\tau,z_I)= \Tr_{RR} \left[ e^{2\pi i \tau (L_0 -c/24)}
e^{-2\pi i \taub (\Lt_0-\ct/24)} e^{2\pi i z_I q^I}(-1)^F\right]~.}
Once we understand this it is straightforward to incorporate potentials
for the rightmoving charges $\qt^I$, if desired.
We'll consider both the Hamiltonian and path integral
approaches, which, as explained in Appendix A, are useful for making
manifest the behavior under spectral flow and modular transformation,
respectively.    In keeping with the Farey tail philosophy \farey, we will explicitly compute the
contribution to the elliptic genus from states below the black hole threshold.
With this in hand, black holes are included by the construction \mc.

\subsec{Hamiltonian approach}

In the Hamiltonian approach we need to enumerate the allowed set of bulk solutions
and their charge assignments.    For the elliptic genus we consider states of the
form (anything, R-ground state), which have $\Lt_0 = {\kt\over 4}$.   There are three
classes of such states: smooth solutions in the effective three dimensional
theory; states coming from Kaluza-Klein reduction of the higher dimensional supergravity
theory; and non-supergravity string/brane states.   Some members of the first class
were discussed above, and we consider the other types of states later.

Just as was done in the CFT approach \adz, it is useful to factorize the dependence on the potentials.
In the gravitational context it is manifest that the stress tensor consists of a
metric part plus  a gauge field part.   Suppose we are given a state
carrying leftmoving charges 
\eqn\agia{
(L_0-{c\over 24}, q^I)=(m,\mu^I)~.
}   
We can apply spectral flow to generate the family of states with charges
\eqn\agi{\eqalign{L_0-{c\over 24} &=   m +\eta_I q^I+k^{IJ}\eta_I\eta_J = m-{1 \over 4}\mu^2 +{1 \over 4}
(\mu+2k\eta)^2 \cr   q^I&= \mu^I +2 k^{IJ} \eta_J~,}}
where we are using the same shorthand notation as in \aeza.   This class of states
will then contribute to the elliptic genus as
\eqn\agj{ \chi(\tau,z_I) = (-1)^F e^{2\pi i \tau (m-{1 \over 4}\mu^2)} \Theta_{\mu,k}(\tau,z_I)~,}
in terms of the $\Theta$-function \aez.
Each such spectral flow orbit has a certain degeneracy from the number distinct
states with these charges.   We call this degeneracy $c(m-{1 \over 4}\mu^2)$, where the
functional dependence is fixed by the spectral flow invariance, and we also include
$(-1)^F$ in the definition.    We can now write down the ``polar" part of the
elliptic genus, that is, the contribution below the black hole threshold:
$m-{1 \over 4}\mu^2 <0$.   We then have
\eqn\agk{ \chi^\prime(\tau,z_I) =
\sum_{m,\mu}\nolimits^{\prime}  c(m-{1 \over 4}\mu^2) \Theta_{\mu,k}(\tau,z_I) e^{2\pi i
(m-{1 \over 4}\mu^2)\tau} ~.}

In the Hamiltonian approach it is easy to write down the polar part of the elliptic
genus in terms of the degeneracies $c(m-{1 \over 4}\mu^2)$.   But the full elliptic
genus also has a contribution from black holes, and these are not easily incorporated
since black holes do not correspond to individual states of the theory.  To incorporate
black holes we need to turn to a Euclidean path integral, as we do now.

\subsec{Path integral approach}

In the path integral approach we sum over bulk solutions with fixed
boundary conditions
\eqn\agl{\chi_{PI}(\tau,z_I) = \sum e^{-S}~.}
The action appearing in \agl\ is
the full string/M-theory effective action reduced to AdS$_3$, though we fortunately do not require
its explicit form to compute the elliptic genus.  In particular, in \agl\ we only
sum over stationary points of $S$ since the fluctuations have already been
incorporated through higher derivative corrections to the action.

The boundary conditions on the metric are that the boundary geometry is a torus
of modular parameter $\tau$.   $z_I$ fix the boundary conditions for the gauge
potentials.  As explained in Appendix A, the relation is, in conformal gauge,
\eqn\agm{A_{I\wb} = {iz_I \over \tau_2}~.}
$A_{Iw}$ is not fixed as a boundary condition.   Since the potential $\zt_I$ is set
to zero in the elliptic genus, we also have the boundary condition
\eqn\agn{ \At_{Iw}=0~.}

Now we turn to the allowed values of $A_{Iw}$ and $\At_{I\wb}$.     The allowed boundary
values of $A_{Iw}$ are determined from the holonomies around the contractible
cycle of the AdS$_3$ geometry.   Recall that when we write $w=\sigma_1+i\sigma_2$
we are taking $\sigma_1$ to be the $2\pi$ periodic spatial angular coordinate.
The corresponding cycle on the boundary torus is contractible in the bulk, and so
any nonzero holonomy must match onto an appropriate source in order to be physical.
The holonomy of a charge $q^I$ particle is
\eqn\agq{ e^{\half i q^I \int\! d\sigma_1 A_{I\sigma_1} } =e^{ \half i q^I \int\!  d\sigma_1 (A_{Iw}
+A_{I\wb})}~.}
Choosing a gauge with constant $A_{Iw}$, we write the allowed values as
\eqn\agr{ A_{Iw} = k_{IJ} \mu^I +2\eta_I -{iz_I \over \tau_2}~,\quad q^I\eta_I \in \IZ~,}
where we have written the charge of the source as $\mu^I$.

In the same way we can determine the allowed values of $\At_{I\wb}$.  In
this case we know that only geometries with $\Lt_0- {\ct \over 24}=0$ contribute
to the elliptic genus, and so we do not include the spectral flowed geometries as
we did above.  Instead, we just have
\eqn\agu{ \At_{I\wb} = \kt_{IJ} \mut^I~.}
%

%We consider the action of a solution carrying charges
%
%\eqn\agp{ L_0 - {c\over 24} = m,\quad q^I,\quad \Lt_0-{\ct\over 24}=0~,\quad \qt^I~.}
%

Given the gauge fields, we know the exact stress tensor \afz\ and also the exact 
currents \afm\ and \afp. We can therefore find the action 
by integrating 
\eqn\ago{\eqalign{ \delta S &= 
\int_{\p AdS}\! d^2x \sqrt{g^{(0)}}\left( {1 \over 2}T^{\alpha\beta} \delta g^{(0)}_{\alpha\beta}+{i\over 2\pi} J^{I\alpha} \delta A_{I\alpha} \right)
\cr &= 
(2\pi)^2 i \left[ -T_{ww} \delta\tau + T_{{\bar w}{\bar w}}\delta {\bar\tau}
+ {\tau_2\over\pi} J^I_w \delta A_{I\wb} + {\tau_2\over\pi} \Jt^I_{\wb} \delta \At_{I w} 
\right]_{\rm const}
~,}}
as in section 4.4.
The result is
\eqn\agt{\eqalign{ S&=  -2\pi i \tau(L_0^{grav}-{c\over
24})+2\pi i \taub (\Lt_0^{grav}-{\ct\over 24})\cr &\quad-{i\pi  \over 2}\left[ \tau A^2_{w}  +\taub
A^2_{\wb}  + 
2\taub A_{w} A_{\wb} \right] +{i\pi  \over 2} \left[ \tau \At^2_{w}  +\taub
\At^2_{\wb} + 2\tau \At_{w} \At_{\wb} \right]~. }}
In verifying that \agt\ satisfies \ago\ one has to take care to consider only variations 
consistent with the equations of motion and the assumed 
boundary conditions. We maintain fixed holonomies by
taking $\delta A_{Iw} = - \delta A_{I\wb}$ and $\delta \At_{Iw} = - \delta\At_{I\wb}$.
Also, the variation of the complex structure must be taken with the gauge field fixed 
in the $z$-coordinates introduced in \ahb. 

The result \agt\ for the action agrees with \ahc\ when the geometry is in the ground 
state where $A_{Iw}\ = - A_{I\wb}$ and  $\At_{Iw}\ = - \At_{I\wb}$, but it is valid also 
more generally in the presence of charged sources.
In fact, it is equivalent to the Hamiltonian result discussed in section 5.1. 
To see this we consider again the charge assignments \agia. Writing
$L_0 = L_0^{grav}  + L_0^{gauge} = L_0^{grav} + {1 \over 4}\mu^2$ (and
analogously for $\Lt_0$) we insert into \agt\ and find
\eqn\agua{ S = -2\pi i \tau(m-{1 \over 4} \mu^2) -{i\pi  \tau\over 2}
(\mu+2k \eta)^2-2\pi i  z_I (\mu^I+2k^{IJ}\eta_J)-{\pi  z^2
\over \tau_2}~. }
Summing over the geometries below the black hole threshold we find
\eqn\agv{\eqalign{ \chi'_{PI}(\tau,z_I) &= \sum\nolimits^{\prime}_{m,\mu} c(m-{1 \over 4}\mu^2) e^{-S}
\cr & = e^{{\pi  z^2
\over \tau_2}}\sum\nolimits^{\prime}_{m,\mu} c(m-{1 \over 4} \mu^2) \Theta_{\mu,k}(\tau,z_I)
e^{2\pi i (m-{1 \over 4} \mu^2)\tau} \cr & = e^{{\pi  z^2
\over \tau_2}}  \chi'(\tau,z_I)~,}}
where $\chi^\prime$ is the Hamiltonian result \agk. As discussed in Appendix A, the overall 
exponential factor is precisely the one we expect.

\subsec{Including black holes}

Black holes are readily included in the path integral approach since they are just
rewritten versions of solutions below the black hole threshold.  Taking a solution below
the black threshold and performing the coordinate transformation $w\rightarrow
{aw + b \over cw +d}$ generates  a black hole.   Using the manifest invariance of the
action under such  coordinate transformations, the contribution of
such a black is then
\eqn\agw{ \chi_{PI}(\tau,z_I) = \chi'_{PI}\left( {a\tau +b \over c\tau +d}, {z_I \over c\tau+d}\right)~. }
On the other hand, from the relation \agv\ between $\chi'_{PI}$ and $\chi'$ we have
\eqn\agx{ \chi'_{PI}\left( {a\tau +b \over c\tau +d}, {z_I \over c\tau+d}\right) =
e^{-2\pi i {c z^2 \over c\tau +d}}e^{{\pi z^2 \over \tau_2}}\chi' \left( {a\tau +b \over c\tau +d}, {z_I \over c\tau+d}\right)~.}
Thus the black hole contribution to $\chi$ is
\eqn\agy{ \chi(\tau,z_I)= e^{-{\pi z^2 \over \tau_2}} \chi_{PI}(\tau,z_I) =
e^{-2\pi i {c z^2 \over c\tau +d}}\chi' \left( {a\tau +b \over c\tau +d}, {z_I \over c\tau+d}\right)~.}

The next step is to sum over all inequivalent black holes to get the complete
elliptic genus.  This means summing over the subgroup of $\Gamma=SL(2,\IZ)$
corresponding to inequivalent black holes or, more precisely, distinct ways of labelling 
the contractible cycle in terms of time and space coordinates.  As explained in \farey\
the inequivalent cycles are parameterized by $\Gamma_\infty \setminus \Gamma$;
so it seems natural to write
\eqn\agz{ \chi(\tau,z_I)=\sum_{\Gamma_\infty \setminus \Gamma}
e^{-2\pi i {c z^2 \over c\tau +d}}\chi' \left( {a\tau +b \over c\tau +d}, {z_I \over c\tau+d}\right)~.}
However, as emphasized in \farey, this cannot be correct since the sum is not convergent.
Instead we should compute not the
elliptic genus but instead its Farey transform, introduced in section 2.
This amounts to  first replacing $\chi'$ by
\eqn\aia{ \hat{\chi}'(\tau,z_I) =\sum\nolimits^{\prime}_{m,\mu} \ct(m-{1 \over 4} \mu^2) \Theta_{\mu,k}(\tau,z_I)
e^{2\pi i (m-{1 \over 4} \mu^2)\tau} }
with $\ct$ defined as in \mda.   We interpret this as the polar part of a weak Jacobi form
of weight $3$ and index $k$.   Instead of \agz\ we therefore write
\eqn\aib{  \hat{\chi}(\tau,z_I)=\sum_{\Gamma_\infty \setminus \Gamma} (c\tau+d)^{-3}
e^{-2\pi i {c z^2 \over c\tau +d}}\hat{\chi}' \left( {a\tau +b \over c\tau +d}, {z_I \over c\tau+d}\right)~.}
In the D1-D5 system this agrees with the Farey transform of the CFT elliptic genus.

\subsec{High temperature behavior}
The high temperature ($\tau_2\rightarrow 0$) behavior of \aib\ is governed by
the free energy of a BPS black hole.  The leading exponential behavior  can be
read off from the term
\eqn\ka{ \left(\matrix{a&b\cr c&d}\right) =\left(\matrix{0&-1\cr
1&0}\right)~,\quad m=0~,\quad \eta_I =0, \quad \mu^I = k \delta^{I0}~,}
which gives
\eqn\kb{ \hat{\chi}(\tau,z_I) \approx  e^{-{2\pi i z^2 \over
\tau} +{2\pi i k z_0 \over \tau }  } ~.}
We can compare with \ahg\ by performing the spectral flow \adb\ $z_0\rightarrow z_0+\half$.
This yields
\eqn\kba{ \ln \hat{\chi}(\tau,z_I)  \approx  {i\pi k \over 2\tau} -{2\pi i z^2 \over \tau}~.}
Noting that this agrees with the holomorphic part of \ahg, we find that the
corresponding entropy is is indeed that of a BPS black hole,
\eqn\kbb{ s= 2\pi \sqrt{{c \over 6} (L_0 - {c \over 24} -{1 \over 4}q^2)  } ~.}
This is just the leading part of the entropy, and is insensitive to the distinction between
the elliptic genus and its Farey-tail transformed version.

\newsec{ Supergravity fluctuations}

To compute the elliptic genus we need to know the spectrum of BPS states,
as described by the coefficients $c( m-{1 \over 4}\mu^2)$.    In this section we
consider the contribution from supergravity states, which are obtained from the
fluctuation spectrum of $10$ or $11$ dimensional supergravity compactified
on AdS$_3$ times some compact space. Our computation generalizes
one given in \GSY.

We will focus on the $(0,4)$ case, corresponding to M-theory on
AdS$_3 \times S^2 \times M_6$, as the $(4,4)$ case is quite well known.  
The $(0,4)$  CFT on the AdS$_3$ boundary describes M5-branes wrapped on 4-cycles
in $M_6$ \MSW\  (the same CFT also describes black rings \bkmicro.)
Up to a spectral flow, supergravity states can carry vanishing charges, $q^I=0$.  These charges
are instead carried by wrapped branes.       So the contribution to the polar part of
the elliptic genus from such supergravity states is
\eqn\aja{ \chi_{sugra} '(\tau,z_I) = \sum\nolimits^{\prime}_m c_{sugra}(m) e^{2\pi i m \tau} ~.}
We will now compute $\chi_{sugra} '(\tau,z_I)$ in order to extract the coefficients
$c_{sugra}(m) $.

\subsec{Spectrum}

It is conventional to compute the elliptic genus in the NS sector, related to the
R sector by the spectral flow \adb.    In the NS sector the elliptic genus
receives contributions from rightmoving chiral primaries obeying $\hbar = \half \qt^0$,
where $\hbar$ is the eigenvalue of $\Lt_0$.   There will be two types supergravity
modes: dynamical modes and ``singletons".   The latter are pure gauge modes
that are nonetheless physical.  We first focus on the dynamical modes.

We work with a 5-dimensional supergravity obtained by compactifying M-theory
on $M_6$, where $M_6$ can be $CY_3$, $K3\times T^2$, or $T^6$ corresponding
to having ${\cal N}=2$, $4$, or $8$ susy.     The 5-dimensional spectrum is written
in the ${\cal N}=2$ language in terms of the number of vectormultiplets $n_V$, hypermultiplets
$n_H$, and gravitino multiplets $n_S$, in addition to the graviton multiplet.
\bigskip
\vbox{
$$\vbox{\offinterlineskip
\hrule height 1.1pt
\halign{&\vrule width 1.1pt#
&\strut\quad#\hfil\quad&
\vrule width 1.1pt#
&\strut\quad#\hfil\quad&
\vrule width 1.1pt#
&\strut\quad#\hfil\quad&
\vrule width 1.1pt#
&\strut\quad#\hfil\quad&
\vrule width 1.1pt#\cr
height3pt
&\omit&
&\omit&
&\omit&
&\omit&
\cr
&\hfil $M_6$ &
&\hfil $n_S$&
&\hfil $n_V $&
&\hfil $n_H$ &
\cr
height3pt
&\omit&
&\omit&
&\omit&
&\omit&
\cr
\noalign{\hrule height 1.1pt}
height3pt
&\omit&
&\omit&
&\omit&
&\omit&
\cr
&\hfil $CY_3$ &
&\hfil $0$&
&\hfil $h^{1,1}-1 $&
&\hfil $2h^{1,2}+2$&
\cr
height3pt
&\omit&
&\omit&
&\omit&
&\omit&
\cr
\noalign{\hrule}
height3pt
&\omit&
&\omit&
&\omit&
&\omit&
\cr
&\hfil $K3\times T^2$ &
&\hfil $2$&
&\hfil $22$&
&\hfil $42$&
\cr
\noalign{\hrule}
height3pt
&\omit&
&\omit&
&\omit&
&\omit&
\cr
&\hfil $T^6$ &
&\hfil $6$&
&\hfil $14$&
&\hfil $14$&
\cr
%\noalign{\hrule}
%height3pt
&\omit&
&\omit&
&\omit&
&\omit&
\cr
}\hrule height 1.1pt
}
$$
}
\centerline{\sl Table 1: 5-dimensional supergravity spectra.}
\def\dfourstates{Table 1}
\def\sugrastates{Table 2}
\bigskip

The chiral primaries form multiplets under the leftmoving $SL(2,\IR)$ symmetry.
In \sugrastates\
we list the spectrum of single particle chiral primaries that are also
primary under the
leftmoving $SL(2,\IR)$; i.e. are annihilated by $L_{1}$.
\bigskip
\vbox{
$$\vbox{\offinterlineskip
\hrule height 1.1pt \halign{&\vrule width 1.1pt#
&\strut\quad#\hfil\quad& \vrule width 1.1pt#
&\strut\quad#\hfil\quad& \vrule width 1.1pt#
&\strut\quad#\hfil\quad& \vrule width 1.1pt#\cr height3pt &\omit&
&\omit& &\omit& \cr &\hfil $s=h-\tilde{h}$& &\hfil degeneracy& &\hfil
range of $\tilde{h}=\half \qt^0$& \cr height3pt &\omit& &\omit& &\omit&
\cr \noalign{\hrule height 1.1pt} height3pt &\omit& &\omit&
&\omit& \cr &\hfil $1/2$& &\hfil $n_H$& &\hfil $1/2,3/2,\ldots$&
\cr height3pt &\omit& &\omit& &\omit& \cr \noalign{\hrule}
height3pt &\omit& &\omit& &\omit& \cr &\hfil $0$& &\hfil $n_V$&
&\hfil $1,2,\ldots$& \cr \noalign{\hrule} height3pt &\omit&
&\omit& &\omit& \cr &\hfil $1$& &\hfil $n_V$& &\hfil $1,2,\ldots$&
\cr \noalign{\hrule} height3pt &\omit& &\omit& &\omit& \cr &\hfil
$-1/2$& &\hfil $n_S$& &\hfil $3/2,5/2,\ldots$& \cr
\noalign{\hrule} height3pt &\omit& &\omit& &\omit& \cr &\hfil
$1/2$& &\hfil $n_S$& &\hfil $3/2,5/2,\ldots$& \cr \noalign{\hrule}
height3pt &\omit& &\omit& &\omit& \cr &\hfil $3/2$& &\hfil $n_S$&
&\hfil $1/2,3/2,\ldots$& \cr \noalign{\hrule} height3pt &\omit&
&\omit& &\omit& \cr &\hfil $-1$& &\hfil $1$& &\hfil $2,3,\ldots$&
\cr \noalign{\hrule} height3pt &\omit& &\omit& &\omit& \cr &\hfil
$0$& &\hfil $1$& &\hfil $2,3,\ldots$& \cr \noalign{\hrule}
height3pt &\omit& &\omit& &\omit& \cr &\hfil $1$& &\hfil $1$&
&\hfil $1,2,\ldots$& \cr \noalign{\hrule} height3pt &\omit&
&\omit& &\omit& \cr &\hfil $2$& &\hfil $1$& &\hfil $1,2,\ldots$&
\cr height3pt &\omit& &\omit& &\omit& \cr }\hrule height 1.1pt }
$$
} \centerline{\sl Table 2: Spectrum of (non-singleton) chiral
primaries for $AdS_3\times S^2\times M_6$
\KutasovZH\foot{The earlier references \deBoerIP\ give incorrect ranges of ${\bar h}$
that differ slightly from these.}.}
\bigskip
Each chiral primary above lies at the bottom of an $SL(2,\IR)$ multiplet
obtained by acting an arbitrary number of times with $L_{-1}$.

\subsec{$NS$ sector elliptic genus}

The contribution from supergravity states  to the $NS$ sector elliptic genus  can be
written
\eqn\ajb{ \chi_{NS}^{sugra}(\tau) = \Tr_{cp} \left[(-1)^{\qt^0} q^{L_0}\right] ~,}
where the trace is over chiral primaries, and $q=e^{2\pi i \tau}$.    As we have said,
the complete spectrum of single particle primaries corresponds to \sugrastates\
and
their $SL(2,\IR)$ descendants.   Multiparticle chiral primaries are obtained by
taking arbitrary tensor products of single particle chiral primaries.

The single particle spectrum starts at $h_{min} = \tilde{h}_{min}+s$.
The contribution of a bosonic tower  $(\qt^0$ even)  is then
\eqn\mg{ \chi_{NS}^{bos}(\tau) =  
\prod_{\ell=0}^\infty\prod_{p=0}^\infty \sum_{m=0}^\infty
q^{m(h_{min}+\ell+p)}=\prod_{\ell=0}^\infty\prod_{p=0}^\infty
 {1 \over 1-q^{(\tilde{h}_{min}+s+\ell+p)}} ~.}
In the above, $m$ stands for the number of particles; $p$ for
acting with
$(L_{-1})^p$; and $\ell$ for $\tilde{h} = \tilde{h}_{min}+\ell$. Now
define $n=\ell+p+1$, so that there are $n$ distinct terms with the
same power of $q$.  Then we can write
\eqn\mh{ \chi_{NS}^{bos}(\tau) =\prod_{n=1}^\infty \left[{1 \over
1- q^{h_{min}-1+n} }\right]^n~. }

The computation for fermions is analogous, and gives
\eqn\mi{ \chi_{NS}^{fer}(\tau) =\prod_{n=1}^\infty \left[{ 1-
q^{h_{min}-1+n} }\right]^n~.}
We can simplify the infinite products using
\eqn\ceba{\eqalign{\prod_{n=1}^\infty (1-q^{1+ n})^n &=M(q)
\prod_{n=1}^\infty{1 \over (1-q^n)}~,\cr \prod_{n=1}^\infty
(1-q^{2+ n})^n &=M(q) \prod_{n=1}^\infty{1 \over (1-q^n)}{1 \over
(1-q^{n+1})} ~,}}
where the McMahon function is defined as
\eqn\ceb{ M(q) = \prod_{n=1}^\infty (1-q^n)^n~.}
The overall power of $M(q)$ will be equal to $n_F- n_B$, the
number of fermonic towers minus the number of bosonic towers. From
the degeneracies in \sugrastates\ and the entries of \dfourstates\
we have
\eqn\ced{ n_F-n_B = n_H +3n_S -2n_V -4= \left\{ \matrix{0&
K3\times T^2 ~{\rm or}~ T^6 \cr 2(h^{1,2}-h^{1,1}) = -{\rm Euler}
& CY_3} \right. }
where ``Euler" denotes the Euler number.

From \sugrastates\ we read off the spectrum of $h_{min}$. For
bosons we have: $n_V+1$ towers with $h_{min}=1$; $(n_V+2)$ towers
with $h_{min}=2$; and $1$ tower with $h_{min}=3$.  For fermions we
have: $(n_S+n_H)$ towers with $h_{min}=1$; and $2n_S$ towers with
$h_{min}=2$.  We then find the supergravity elliptic genus to be
\eqn\ci{ \chi^{sugra}_{NS}(\tau) = M(q)^{-{\rm
Euler}}\prod_{n=1}^\infty (1-q^n)^{n_v+3-2n_s}(1-q^{n+1})~. }

\subsec{Spectrum of gauge fields and their Chern-Simons couplings}

We now discuss the three dimensional spectrum of gauge fields relevant to 
compactifying M-theory on AdS$_3 \times S^2 \times M_6$, with $M_6$ 
being $T^6$, $K3\times T^2$ or $CY_3$.  Besides being an important general property of these theories,
the precise spectrum will needed in the next subsection when we work out the singleton contribution 
to the elliptic genus. 

In five dimensions there are a total of $n_V+2n_S+ 1$ gauge fields, the $+1$ coming from
the gauge field in the graviton multiplet.   The action for these gauge fields includes
the Chern-Simons coupling $\int\! C^{IJK} A_I \wedge F_J \wedge F_K$, where $C^{IJK}$ is the intersection form on $M_6$. 
To pass to the three dimensional description we reduce on $S^2$ in the presence of
magnetic flux,  $p_I = {1 \over 2\pi  }\int_{S^2} F_I$.     The three dimensional 
Chern-Simons term is therefore $\int\!  C^{IJK}p_K A_I \wedge F_J$.  
In addition to these $U(1)$ Chern-Simons terms we also have   the $SU(2)$  Chern-Simons term for the  $S^2$ Kaluza-Klein gauge fields.

We now discuss each of the choices of $M_6$ in turn.

\vskip.2cm \noindent  $ {\bf T^6}  $

M-theory on $T^6$ has $27$ gauge fields transforming in the ${\bf 27}$ of the
$E_{6(6)}$ duality group.   The Chern-Simons coupling $C^{IJK}$ is the 
$E_6$ cubic invariant.   We consider the case where we turn on a single
magnetic charge.   It is then convenient to decompose the ${\bf 27}$ under
an $SO(5,5)$ subgroup as ${\bf 27 \rightarrow 16 + 10+1}$, and to turn on
the singlet.    The cubic invariant decomposes as ${\bf 27^3 \rightarrow 
1 \cdot 10 \cdot 10   + 16 \cdot 16 \cdot 10}$.    The three dimensional 
Chern-Simons term is then $\int\! g^{IJ}A_I \wedge F_J$ where $g^{IJ} = {\rm diag} ( (+1)^5, (-1)^5)$ is the $SO(5,5)$ invariant quadratic
form.   So 
we find $5$ leftmoving and $5$ rightmoving $U(1)$ currents, in addition to the 
rightmoving $SU(2)$ R-currents.  

This result has a simple interpretation if we take, by U-duality, the charge
to correspond to a wrapped  fundamental string in IIA compactified on $T^5$.    
The $(5,5)$ currents correspond to momentum/winding charges on the 
$5$ transverse compact dimensions.   In addition, on the worldsheet there 
are three additional left and right moving currents corresponding to translations
in the noncompact directions.  

In comparing the AdS and worldsheet descriptions we note two basic facts.
First, although this charge configuration is expected on general grounds 
to yield a near horizon AdS$_3 \times S^2 \times T^6$ geometry, this is yet to be
shown.  Just keeping the two derivative terms in the spacetime action leads to 
a naked singularity.   Furthermore, in this case there are no $R^2$ corrections
in the action that might stabilize the geometry in analogy with other examples.
Establishing the existence of a stabilized geometry requires a better understanding of $R^4$ type terms than is
currently available.      Second, while the $(5,5)$ currents are found to match
in the AdS and worldsheet descriptions, we note that the translation currents
are absent on the AdS side.   On the worldsheet there are $3$ left and $3$ rightmoving
currents corresponding to motion in the transverse noncompact directions.
These center of mass degrees of freedom have
been ``factored out" in passing to the near horizon geometry.

\vskip.2cm \noindent $ {\bf K3\times T^2}$

M-theory on $K3 \times T^2$ has  duality group  $SO(21,5) \times SO(1,1)$, 
as is readily understood by dualizing to the heterotic string on $T^5$.  
There are $27$ gauge fields transforming as ${\bf 26 + 1}$ under $SO(21,5) $.  
By direct computation on the heterotic side one finds that the Chern-Simons
coupling in five dimensions is of the form $\int g^{IJ} A_I \wedge F_J \wedge F_{27}$,
where $g^{IJ}$ is the $SO(21,5)$ invariant quadratic form, and $F_{27}$ refers
to the $SO(21,5)$ singlet gauge field.   Magnetic charge with respect to the singlet
corresponds to an M5-brane wrapped on $K3$, dual to a fundamental string
on the heterotic side.  We consider turning on only this magnetic charge,
Again, in the two derivative approximation this yields a singular spacetime solution,
but in this case it has been shown explicitly \CardosoFP\  that $R^2$ corrections stabilize the 
geometry into a smooth AdS$_3 \times S^2 \times K3 \times T^2$.    Reducing the
five dimensional Chern-Simons term to three dimensions yields $\int g^{IJ} A_I \wedge F_J$.
We thus find that the spectrum of currents matches up with the worldsheet structure
of the heterotic string.  Again, the translational currents are absent in the AdS
picture; adding them in yields the $(24,8)$ spectrum of currents corresponding to
the transverse modes of the heterotic string. 

\vskip.2cm \noindent $ {\bf CY_3}$

There are $n_V +1 = h^{1,1}$ gauge fields with five dimensional Chern-Simons 
coupling $\int C^{IJK} A_I \wedge F_J \wedge F_K$ where $C^{IJK}$ is the
intersection form of the $CY_3$.    
Reduction to three dimensions give a signature $(n_V,1)$ Chern-Simons term,
as follows from the Hodge index theorem (see \MSW).

\subsec{Including singletons}

Singleton modes are pure gauge configurations that are nonetheless
physical in the presence of the AdS$_3$ boundary.   To see why, consider
the case of a $U(1)$ gauge field with Chern-Simons term.  The
configuration $A_w = \p_w \Lambda(w)$ is formally pure gauge, but from \afr\
it carries the nonzero energy $T_{ww} = {k \over 8\pi} (\p_w \Lambda)^2$,
and hence is physical. This is possible because the true gauge transformations
must vanish at the boundary and it is only those that leave the stress tensor invariant.
The singleton states are described in the CFT as
$J_{-1}|0\rangle$, where $J$ is the current corresponding to $A$.   We also have
the $SL(2,\IR)$ descendants of these states.

A similar story holds for singletons associated with diffeomorphisms that are
nonvanishing at the boundary.     These correspond to the states
$L_{-2}|0\rangle$ and $SL(2,\IR)$ descendants thereof.   The explicit form of the diffeomorphisms is given in \BrownNW.

We can now work out the contribution of the singletons to the elliptic genus
of the $(0,4)$ theory.   The computation is the same as in \mg\ except without
the product over harmonics $\ell$.   The leftmoving currents have
$h_{min}=1$, and the stress tensor has $h_{min}=2$.    If there
are $n_L$ leftmoving currents then the contribution of singletons is
\eqn\mj{\chi_{NS}^{sing} = \prod_{n=1}^\infty {1 \over
(1-q^n)^{n_L}}{1 \over ( 1-q^{n+1}) } }

Now, we found in the previous subsection that 
\eqn\mk{ n_L= \left\{ \matrix{5 & T^6 \cr 21 & K3\times T^2 \cr
n_V & CY_3 } \right.~.  }
We find the full results by multiplying \ci\ and \mj:
\eqn\mka{ \chi_{NS} =\chi_{NS}^{sugra} \chi_{NS}^{sing}=  \left\{ \matrix{1 & T^6 \cr &\cr  1 &
K3\times T^2 \cr & \cr M(q)^{-\chi_E}\prod_{n=1}^\infty
(1-q^n)^{3} & CY_3 } \right.  }
We find that in the $T^6$ and $K3\times T^2$ cases the singletons
precisely cancel the dynamical contribution \ci. For the CY$_3$ the dependence
on $n_V$ cancelled.  Note that these conclusion are a result of cancellations between propagating
states from \sugrastates\ and the singletons.

%This makes sense: we expect
%that the entire elliptic genus should cancel in the $T^6$ case,
%and in the $K3\times T^2$ case  should come from M2-branes wrapped
%on $T^2$. I'm not sure what the significance of the $(1-q^n)^3$
%factor in the $CY_3$ case is.

\subsec{R sector}

As far as the supergravity fluctuations are concerned, the NS and R
sector elliptic genera are identical in the $(0,4)$ case.  In general we
have the relation \adb, but in the $(0,4)$ case there is no $z_0$ since
there is no leftmoving R-symmetry.   So in \mka\ we can trivially replace
NS by R.

\newsec{Contribution from wrapped branes}

The final ingredient in the computation of the elliptic genus is the contribution
from wrapped branes.  In \GSY\ it was shown that this computation is equivalent
to the Gopakumar-Vafa derivation \GopakumarII\  of the topological string partition function from 
M-theory.   The same argument applies here, and so we can be brief.  

The elliptic genus receives contribution from M2 branes and antibranes wrapping 
2-cycles in $M_6$ and carrying angular momentum on $S^2$ \GSY.   In the general
$CY_3$ case an explicit result requires a determination of the BPS spectrum of
M2-branes.   In the $T^6$ and $K3\times T^2$ examples emphasized here there are
drastic simplifications, as discussed in \KatzXQ.   In the $T^6$ case there is no
contribution at all since the M2-branes are in sufficiently large multiplets to mutually
cancel.   In the $K3\times T^2$ case there is a similar cancellation except for M2-branes
wrapping $T^2$.  These M2-branes precisely reproduce the known correction to the
topological string partition function on $K3\times T^2$.   From the standpoint of section
6.3, these M2-branes are states electrically charged with respect to $F_{27}$, which
we recall was the one special gauge field not appearing in the Chern-Simons term.

\newsec{Discussion and open questions}

Let us review what has been achieved.   We have shown, following the ideas in \farey, how to systematically 
compute the elliptic
genus (or rather, its Farey tail transform) of string/M-theory on AdS$_3$ using supergravity.   
What makes this possible is that the long distance theory on AdS$_3$ is topological,  allowing for
the exact determination of currents and stress tensors.   The currents and stress tensor 
can be integrated to find the action, and then summing over the space of solutions yields the
elliptic genus.   What distinguishes different examples from one another is the precise spectrum of 
currents and the spectrum of BPS states supporting nontrivial holonomies.   We also noted that the same
approach can be employed in the computation of the full partition function, but here one would need the
full spectrum of states and not just its BPS sector, and so explicit results are not possible. 

To be more specific, the path integral evaluation of the Farey tail transformed elliptic genus led to the
expressions \aia-\aib.     The theta function arises from summing over the allowed class of gauge
fields, where it's important that we have correctly included all boundary terms in the action.  The 
term in the exponential of \aia\ is the contribution of the purely gravitational part of the action.  
Finally, the coefficients $\tilde{c}(m-{1 \over 4}\mu^2)$ encode the spectrum of BPS states, or more
accurately their orbits under spectral flow.     Once these are known we  can feed \aia\ into \aib\
to complete the computation.    

The $\tilde{c}(m-{1 \over 4}\mu^2)$ are computed from the spectrum of supergravity fluctuations
and wrapped branes.   As discussed  in section 6, the contribution from supergravity fluctuations
is extracted from the Kaluza-Klein spectrum for the compactification of interest.   For the contribution
from wrapped branes we relied on the observation of \GSY\ that this is equivalent to the 
Gopakumar-Vafa computation of the topological string free energy.

Of course, one natural question is why we should be computing the Farey tail transform of the elliptic
genus, rather than the elliptic genus itself.    Recall that the latter cannot quite be extracted from the
former, since states with $m-{1 \over 4} \mu^2=0$ are projected out by the transform.  In \farey\ it was
shown that this procedure is necessary in order for the CFT result to take a natural form in terms
of buk geometries.   It would be nice to have a better understanding of this from the bulk point of view

Our considerations have been from the bulk point of view of the AdS$_3$/CFT$_2$ correspondence.
An exact expression for the  CFT elliptic genus  in the $(4,4)$ case was the starting point for \farey,
and corresponding study of the $(0,4)$ case are the subject of recent investigations 
\refs{\strings,\GaiottoWM}.  
This should lead to a more detailed understanding of the AdS/CFT correspondence, and its
connection with topological strings. 

%\vskip 3.0cm
\bigskip
\noindent {\bf Acknowledgments:} \medskip \noindent We thank J. de Boer,
I. Dolgachev, J. Harvey, 
and E. Verlinde for discussions.
The work of PK is supported in part by the NSF grant
PHY-00-99590. The work of FL is supported  by DoE under grant
DE-FG02-95ER40899.

\appendix{A}{Modular properties of the charged boson partition function }
In this appendix we determine the modular transformation of the partition function
for a single
charged boson by comparing the canonical and path integral formulations.
This illustrates some general features.

Consider a free compact boson of radius $2\pi R$. We use the
conventions of \PolchinskiRQ\ and set $\alpha'=1$. We define the
partition function
\eqn\ga{ Z(\tau,z,\zt) = (q\qb)^{-1/ 24} \Tr\left[ q^{L_0}
\qb^{\Lt_0} e^{2\pi i z p_L} e^{2\pi i \zt p_R} \right]~,}
with
\eqn\gb{\eqalign{ L_0 &= {p_L^2 \over 4} +L_0^{osc}~, \quad  \Lt_0
= {p_R^2 \over 4} +\Lt_0^{osc} \cr p_L &= {n \over R}+wR~,\quad
p_R = {n\over R}-wR~.}}
The partition function obeys the modular transformation rule
\eqn\gh{ Z({a\tau + b \over c\tau +d},{z\over c\tau +d},{\zt\over c\taub +d})
=e^{ { 2\pi icz^2 \over c \tau+d}}e^{ -{2\pi i c\zt^2 \over c\taub+d} }
Z(\tau,z,\zt)~,}
as is readily verified by direct computation.  \gh\ is to be compared with \ab.

To explain the origin of the exponential prefactors in \gh\ we pass to a path
integral formulation.   We consider
\eqn\gj{ Z_{PI}(\tau,A) = \int\! {\cal D}X e^{-S} }
with
\eqn\gk{ S = {1 \over 2\pi}\int_{T^2}d^2\sigma \sqrt{g}\left[ {1
\over 2}g^{ij}\p_i X \p_j X  -A^i \p_i X \right] }
and $A^i =$ constant.  To relate potentials appearing in \ga\ and
\gk, we use the standard expression for the charges
\eqn\gka{ p_L = 2 \oint {dw \over 2\pi i}i\p_w X~,\quad p_R = -2
\oint {dw \over 2\pi i}i\p_{\wb} X~,}
and then equate the charge dependent phases in the two versions.
This yields
\eqn\gn{ z= -i\tau_2 A_{\wb}~,\quad \zt=i\tau_2 \At_w~.}
We denoted the holomorphic part of the gauge field $\At_w$ because,
in the body of the paper,  this component arises from an independent bulk 
1-form $\At$. 

In the path integral formulation a modular transformation is a
coordinate transformation combined with a Weyl transformation, and
so it is manifest that
\eqn\gha{ Z_{PI}({a\tau + b \over c\tau +d},{z\over c\tau
+d},{\zt\over c\taub +d}) = Z_{PI}(\tau,z,\zt)~,}
where the transformation of $z$ and $\zt$ just expresses the
coordinate transformation.

What then is the relation between $Z_{PI}$ and $Z$?  To find this we
just carry out the usual steps that relate Hamiltonian and path
integral expression: $\int \! {\cal D}X e^{-S} = \Tr e^{-\beta H}$.
The only point to be aware of is that the Hamiltonian corresponding
to the action \gk\ is not the factor appearing in the exponential of
\ga, but differs from this by a contribution quadratic in the
potentials.  In particular, we find
\eqn\ghb{ Z_{PI}(\tau,z,\zt) = e^{{\pi (z+\zt)^2 \over
\tau_2}}Z(\tau,z,\zt)~.}
Combining \gha\ and \ghb\ we see that the modular transformation law
of $Z$ must be such to precisely offset that of  $e^{{\pi (z+\zt)^2
\over \tau_2}}$.  This is what \gh\ does.

Let us summarize the lessons just learned as applied to a more
general setting. In the canonical form \ga\ properties such as
spectral flow are manifest.  In the path integral form \gk\ the
modular transformation law is manifest.  By relating the two
versions we can understand both properties.  Furthermore, the
analysis we performed is essentially completely general, in that
given an arbitrary CFT we can always realize the $U(1)$ current
algebra in terms of free bosons.  This is the way we derive \ab\ and
\ac, for example.

\appendix{B}{Conventions}
In this appendix we summarize our conventions.

We work in Euclidean signature with
$\epsilon_{12...}=\sqrt{g}$ and  $\epsilon^{12...}=1/\sqrt{g}$.
The rule for integrating differential forms is
\eqn\fc{ \alpha = \alpha(x) dx^1\wedge \cdots \wedge dx^d \quad
\rightarrow \quad \int \! \alpha = \int \! d^d x  \, \alpha(x)~.}
The star operation in d-dimensions is
\eqn\faa{ {\star} dx^\mu \wedge \ldots \wedge dx^{\mu_p} = {1
\over (d-p)!} \epsilon^{\mu_1 \ldots \mu_p}_{~~~~~~~\mu_{p+1}
\ldots \mu_d} dx^{\mu_{p+1}} \wedge \ldots \wedge dx^{\mu_d}~.}
The components of a p-form are defined as
\eqn\fa{A^{(p)}= {1 \over p} A^{(p)}_{\mu_1 \ldots \mu_p} dx^{\mu_1}\wedge  \ldots  \wedge dx^{\mu_p}  ~,}
so for a 1-form in $d=2$:
\eqn\fb{ {\star}  A^{(1)} = \epsilon^\mu_{~\nu} A^{(1)}_\mu
dx^\nu, \quad {\star}A^{(1)}_\mu = -\epsilon_\mu^{~\nu}
A^{(1)}_\nu, \quad {\star}{\star}A^{(1)}=-A^{(1)}~.}
Useful  wedge products are
\eqn\fb{\eqalign{ A^{(1)}\wedge B^{(1)}& =  A^{(1)}_\mu
B^{(1)}_\nu dx^\mu \wedge dx^\nu= \sqrt{g}\epsilon^{\mu\nu}
A^{(1)}_\mu B^{(1)}_\nu dx^1 \wedge dx^2~, \cr A^{(1)}\wedge
A^{(2)} &= \half A^{(1)}_\mu A^{(2)}_{\nu\sigma} dx^\mu \wedge
dx^\nu \wedge dx^\sigma = \half \sqrt{g}
\epsilon^{\mu\nu\sigma}A^{(1)}_\mu A^{(2)}_{\nu\sigma} dx^1 \wedge
dx^2 \wedge dx^3~,\cr {\star}A^{(1)}\wedge B^{(1)}& =
-\epsilon_\mu^{~\nu} A_\nu^{(1)} B^{(1)}_\sigma dx^\mu \wedge
dx^\sigma = - \sqrt{g} g^{\mu\nu}A_\mu^{(1)} B^{(1)}_\nu dx^1
\wedge dx^2~,}}
where the last formula applies in $d=2$.  From this it follows
that
\eqn\fba{{\star}A^{(1)}\wedge B^{(1)} = {\star}B^{(1)}\wedge
A^{(1)} =-A^{(1)}\wedge {\star}B^{(1)}~.}

Complex coordinates on the boundary are defined as $w=\sigma_1 +i\sigma_2$, where
$\sigma_2$ is Euclidean time.  We then have the components of a vector
\eqn\fca{\eqalign{ v_w & = {v_1-iv_2 \over 2}~,\quad v_{\wb} = {v_1 +iv_2 \over 2}~, \cr
v_1 & = v_w + v_{\wb}~,\quad v_2 = i(v_w -v_{\wb})~.}}
Also useful are Hodge stars
\eqn\fd{ \star dw =-i dw~,\quad \star d\wb = id\wb~.}

\listrefs

\end